\documentclass[a4paper,fleqn,usenatbib]{mnras}

\usepackage{newtxtext,newtxmath}
\usepackage[T1]{fontenc}
\usepackage{ae,aecompl}

\usepackage{graphicx}	% Including figure files
\usepackage{amsmath}	% Advanced maths commands
\usepackage{amssymb}	% Extra maths symbols
\usepackage{changepage} % Added by RA to fit Table 2, which is very large

\usepackage{subcaption}
\newcommand{\Chandra}{${\it Chandra}$}
\newcommand{\Spitzer}{${\it Spitzer}$}

\newcommand\nodata{ ~$\cdots$~ }%

\title[Multiwavelength survey of Sculptor Dwarf X-ray sources]{Multiwavelength survey of X-ray sources in the Sculptor Dwarf Spheroidal Galaxy}

\author[R. M. Arnason et al.]{
R. M. Arnason,$^{1}$\thanks{E-mail: rarnaso@uwo.ca}
P. Barmby$^{1}$,
A. Bahramian$^{2,3}$,
T. J. Maccarone$^{4}$,
S. E. Zepf$^{3}$
\\
$^{1}$Department of Physics and Astronomy, University of Western Ontario, 1151 Richmond Street, London, ON N6A 3K7, Canada\\
$^{2}$International Centre for Radio Astronomy Research, Curtin University, GPO Box U1987, Perth, WA 6845, Australia \\
$^{3}$Department of Physics and Astronomy, Michigan State University, East Lansing, MI 48824, USA\\
$^{4}$Department of Physics, Texas Tech University, Box 41051, Lubbock, TX, 79409-1051, USA
}

\date{Accepted XXX. Received YYY; in original form ZZZ}

\pubyear{2019}

\begin{document}
\label{firstpage}
\pagerange{\pageref{firstpage}--\pageref{lastpage}}
\maketitle

\begin{abstract}
We present an unprecedented, deep study of the primordial low-mass X-ray binary population in an isolated, lower-metallicity environment.
We perform followup observations of previously-identified X-ray binary candidates in the Sculptor Dwarf Galaxy by combining a second \Chandra\ observation with \Spitzer\ and Gemini photometry, as well as Gemini spectroscopy of selected targets.
Of the original nine bright X-ray sources identified, we are able to classify all but one as
quasars, active galactic nuclei, or
background galaxies.
We further discover four new X-ray sources in the second-epoch \Chandra\ observation. Three of these new sources are background sources and one is a foreground flaring star.
We have found that Sculptor is effectively devoid of X-ray sources above a few $10^{34}$ erg~s$^{-1}$.
If Sculptor is able to retain primordial binaries at a similar rate to globular clusters, this implies that bright X-ray binaries observed in globular clusters in the present epoch are all formed dynamically.
\end{abstract}

\begin{keywords}
X-rays:binaries -- X-rays:galaxies -- galaxies:individual:Sculptor dwarf spheroidal
\end{keywords}

%%%%%%%%%%%%%%%%%%%%%%%%%%%%%%%%%%%%%%%%%%%%%%%%%%

%%%%%%%%%%%%%%%%% BODY OF PAPER %%%%%%%%%%%%%%%%%%

\section{Introduction}

X-ray binaries (XRBs) are valuable tracers of star formation history and compact object physics in galaxies.
In old stellar populations, the brightest X-ray sources are low-mass X-ray binaries (LMXBs),  where a black hole or neutron star accretes from a low-mass companion overflowing its Roche Lobe.
These sources have X-ray luminosities $L_{X} > 10^{34}$ erg~s$^{-1}$ \citep{Psaltis04a,Heinke10a}.
Their X-ray spectra include a soft, blackbody-like thermal component attributed to the accretion disk, and a harder non-thermal component attributed to a hot corona of material around the compact object \citep{Done07a}.
At ultraviolet through visible wavelengths, XRB emission combines light from the companion star, a blue continuum of reprocessed light from the accretion disk, and bright emission lines due to atoms excited by the accretion flow \citep{vanParadijs95a,Heinke10a}.
Many X-ray sources are observed to experience transient behaviour, where bright outbursts occur between long periods of quiescence.
This transient behaviour is typically attributed to instability in the accretion disk due to varying viscosities from hydrogen ionization \citep{Lasota01a}.

\subsection{XRB Production and Lifetime}

X-ray binaries can be formed in stellar populations through two main pathways.
Primordial formation involves progenitor binary systems that survive the supernova event of the accretor.
It is generally expected that XRB production should peak in a population roughly 1~Gyr after the peak of star formation \citep{White98a}.
Dwarf irregular galaxies often have high specific star formation rates and contain large numbers of high-mass X-ray binaries \citep[e.g. IC~10;][]{Laycock2017a,Laycock2017b}.
In areas of high stellar density, such as globular clusters (GCs), XRBs can be formed through dynamical encounters: dynamical considerations show that the XRB formation rate is proportional to the square of the stellar density $\rho^2$ \citep{Bahramian13a,Verbunt87a}.
A compact object may be swapped into a binary system in an exchange encounter \citep{Hills91a}, or a compact object passing near two stars may remove enough energy from the motions of the two stars to cause them to be bound into a binary configuration \citep{Verbunt87b,Verbunt06a}.
If stellar density is high enough, new binary systems can be created by physical collisions between stars \citep{Fabian75a,Ivanova06a,Ivanova08a}.
The importance of dynamical encounters in creating XRBs is shown by the fact that they are roughly 100 times more abundant per unit mass in GCs compared to the Galactic field \citep{Clark75a,Verbunt06a}.

The efficiency of globular clusters at producing XRBs led to the suggestion that many, or most XRBs are formed within globular clusters and then captured by their host galaxy through dynamical processes \citep{Grindlay88a,White02a}.
A counter-argument was provided by
investigations of the radial distribution of LMXBs, which show they tend to trace galaxies' stellar mass rather than their GC distributions \citep{Gilfanov04a}.
The advent of gravitational-wave observations has focused attention on simulations of black hole XRBs in GCs; recent results explore BH-XRB production and subsequent ejection in detail \citep{Giesler2018,Kremer2018}.

Once formed, how long does an LMXB system last?
Steady-state accretion would require that a 1~M$_{\sun}$ low-mass X-ray binary (LMXB) companion should be consumed by its accretor within roughly 100~Myr.
However, this expected LMXB lifetime can be altered by a number of factors.
First, many LMXBs are not persistently bright and are observed to undergo long periods of quiescence \citep{Piro02a}.
If LMXBs are quiescent in excess of 75\% of their accretion cycle, there should be a very large number of quiescent LMXBs (qLMXBs) for each bright one.
Investigations of the populations of XRBs in globular clusters suggests that this is the case, with $\sim$10 persistent, bright XRBs over all Galactic globular clusters, and at least $\sim$50 quiescent or transient XRBs found in each of the largest clusters \citep{Fragos09a,Bahramian14a,Heinke05a,Heinke06a}.
Delays in mass transfer can also lengthen an XRB's lifetime.
Ultracompact systems and systems which began as intermediate mass X-ray binaries but evolved to a lower companion mass through mass transfer are both expected to have longer lifetimes \citep{Bildsten04a,Podsiadlowski02a}.
Additionally, some currently existing persistently bright systems may have had no mass transfer at earlier epochs until the donor had evolved on the subgiant branch to a point that it began filling its Roche lobe \citep{Revnivtsev11a}.

\subsection{XRB Populations in Dwarf Galaxies}
Dwarf galaxies are potentially useful in understanding both XRB formation pathways.
Globular clusters associated with dwarf galaxies could be a source of X-ray binaries for their host galaxies.
For example, the Sagittarius Dwarf Spheroidal (Sgr dSph) is in the process of being tidally disrupted by the Milky Way \citep{Sbordone07a}.
Palomar 12 is believed to have been tidally stripped from Sgr dSph and is part of the associated tidal stream \citep{Cohen04a};
M54 is generally thought to be a GC that has sunk to Sgr dSph's core through dynamical friction \citep{Siegel07a}.
In principle, dwarf spheroidal galaxies could also constrain the long-term evolution of primordial XRBs, especially in a low-metallicity environment.
Dwarf spheroidal galaxies tend to have relatively simple star formation histories, often characterized by a brief early star formation event  with a small tail of more recent star formation \citep{Monkiewicz99a}.
Because of their low stellar densities (roughly two orders of magnitude lower than those of globular clusters), dwarf galaxies are unlikely to have any new XRBs formed through dynamical interactions.
The presence of LMXBs can help independently constrain the size of dwarf galaxy dark matter halos, since the supernova kick given to an LMXB progenitor will generally exceed the velocity dispersion of stars in the population \citep{Dehnen06a} and could in principle eject the system from a dwarf galaxy.
However, a dark matter halo could permit a dwarf galaxy to retain XRBs that it would have otherwise lost to the natal kicks or its host galaxy's tidal field \citep{Podsiadlowski05a,Dehnen06a}.

Detection and characterization of XRBs in the Milky Way's dwarf spheroidal galaxies is enhanced compared to that in its globular clusters.
The galaxies' lower central stellar densities mean that multi-wavelength studies suffer less from crowding even though they are on average an order of magnitude more distant than the GCs.
While some dwarf galaxies do contain central black holes,  the typically low radiative efficiency of the BH does not overwhelm the XRB signal \citep{Nucita2017}.
Although not the focus of this paper, the proximity of the many satellite dwarf galaxies to the Milky Way means that transient LMXBs can potentially be discovered in quiescence (below $10^{33}$ erg~s$^{-1}$), if observations are sufficiently deep.

Since the advent of \Chandra\, a number of studies have been conducted to look for LMXB candidates in dwarf galaxies.
The Carina, Sagittarius, Fornax, Leo I, Ursa Major II, and Ursa Minor dwarf galaxies have all been targeted with either \Chandra\ or \textit{XMM-Newton} \citep{Ramsay06a,Nucita13a,Manni15a}.
However, these studies have generally either had low spatial resolution or no multiwavelength counterpart matching.
Without multiwavelength counterpart matching, typically the analysis can only compare integrated properties of the X-ray population with the expected population of background sources.
More recently, two separate studies of the X-ray sources in Draco dSph were conducted by \cite{Sonbas16a} and \cite{Saeedi16a} using \textit{XMM-Newton}.
These surveys both identified four LMXB candidates through matching to the Sloan Digital Sky Survey (SDSS), Wide-Field Infrared Survey Explorer, Two Micron All-Sky Survey, and other visible and infrared surveys.

\subsection{Sculptor Dwarf Spheroidal Galaxy}
The Sculptor Dwarf Spheroidal Galaxy%
\footnote{Not to be confused with the Sculptor dwarf irregular galaxy, which is not a Milky Way satellite but a member of the more distant Sculptor group of galaxies.}
(hereafter Sculptor) is ideal as a low-density counterpart to globular clusters.
Sculptor is one of the closest dwarf galaxies, and is at a favourable Galactic latitude outside the plane of the Milky Way \citep{McConnachie12a}.
It lacks globular clusters of its own, and has a relatively simple star formation history.
Colour-magnitude diagram analyses have suggested that Sculptor has a predominantly ancient stellar population and a smaller population of intermediate age stars  \citep{Monkiewicz99a,Dolphin02a,Tolstoy03a}.
A more recent analysis has shown that this intermediate age population tends to be concentrated towards Sculptor's core, and can be described with a simple, smoothly decreasing star formation rate ending around 7 Gyr ago \citep{deBoer12a}.

\cite{Maccarone05a} surveyed Sculptor using 21 6--ks \Chandra\ pointings, and identified 9 X-ray sources with sufficient counts to accurately identify position and search for optical counterparts.
These sources were matched to the optical catalogue of \cite{Schweitzer95a}, and one source was ruled out as a background galaxy, while five were identified as LMXB candidates, with X-ray luminosity $>10^{33}$~erg~s$^{-1}$.
This result was surprising, as a galaxy of this size would not be expected to have such a large LMXB population: more recently for a sample of Virgo cluster and field dwarfs, \cite{Papa2016} found about 1 bright LMXB per $10^9$~M$_{\odot}$ of stars.
In this paper we re-investigate the nature of bright X-ray binary candidates in Sculptor.
We combine new non-simultaneous Gemini/GMOS imaging (2016) and spectroscopy (2008), along with \Spitzer\ photometry (2008) and a second epoch of \Chandra\ imaging.
Gemini spectroscopy permits us to directly separate contaminating active galactic nuclei (AGN) and foreground stars from objects within Sculptor, while \Spitzer\ and Gemini photometry allow us to look for long-wavelength counterparts to X-ray sources that are associated with Sculptor's population.
We use the mid-infrared AGN selection wedge of \cite{Stern05a} to indicate whether an individual X-ray source is likely to be an AGN.

For this analysis, we assume that Sculptor is at a distance of 86~kpc, with a heliocentric velocity v$_{\sun} = 111.4$~km~s$^{-1}$, and a spectroscopic redshift $z = 0.000372$ \citep{McConnachie12a}.
We use the IRSA dust maps and the \cite{SchaflyFinkbeiner11a} conversion to obtain a value of A$_{V} = 0.0484$ for the Milky Way foreground extinction in the direction of Sculptor.
We further use the relations of \cite{Bahramian15a} and \cite{Foight16a} to obtain a foreground column density of $N_{\rm H} = 9.0 \times 10^{19}$~cm$^{−2}$ for Sculptor.
We assume that Sculptor has negligible internal gas and dust:
21~cm radio observations in the direction of Sculptor detect a total \ion{H}{i} mass of $\sim2\times10^{4}$~M$_{\sun}$ \citep{Bouchard03}, low amongst Local Group dwarfs for which \ion{H}{i} measurements exist \citep{McConnachie12a}.

\section{Data}

\subsection{\Chandra\ Data Reduction}
\label{sec:chandrared}
There are two sets of \Chandra\ observations of Sculptor in the Chandra Data Archive.

\begin{figure*}
    \centering
    \includegraphics[width=\textwidth]{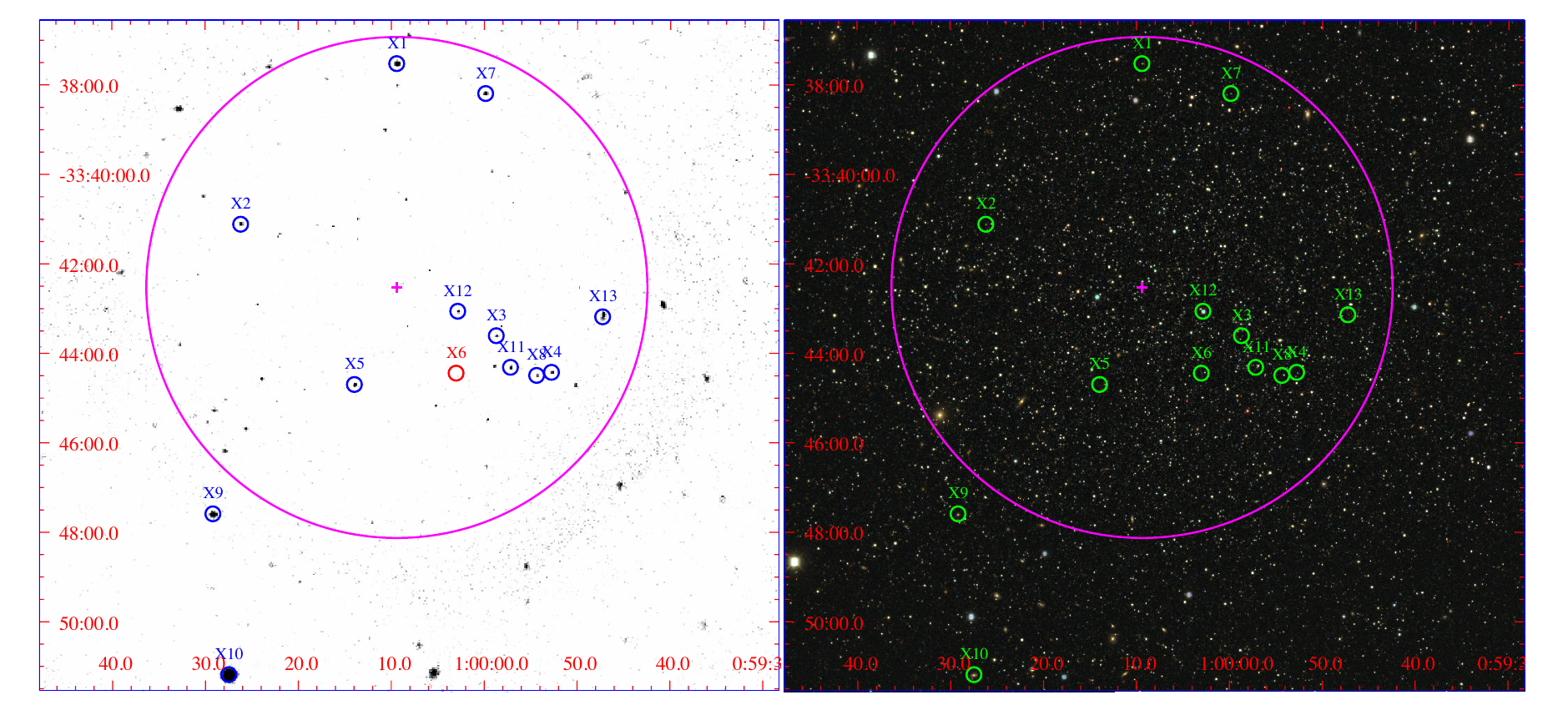}
    \caption{Left: \textit{Chandra} image of Sculptor. Blue and red circles mark X-ray sources that were investigated in this study in detail. Positions are taken from the \texttt{wavdetect} source list described in Section \ref{sec:chandrared}, except for source X6, which is taken from \protect\cite{Maccarone05a} since it was not detected by our analysis.
    Right: Dark Energy Survey combined $g$ (blue),$r$ (green), $i$ (red) image of Sculptor. Green circles correspond to the blue circles in the left panel.
    In both panels, the cross marks the optical centre of the galaxy and the large pink circle marks the core radius $r_{c} = 145$~pc.
    }
    \label{xray_des_image}
\end{figure*}

We summarize these observations in Table \ref{xrayobs}.
For this analysis, we considered both a series of 21 6-ks observations first studied by \cite{Maccarone05a} as well as a new, longer, observation (ObsID: 9555)
The new observation was made with the Advanced CCD Imaging Spectrometer (ACIS-S) in faint mode with an effective frame time of 3.1~s.
We reduced the observations using CALDB version 4.7.2, the August 2008 time-dependent gain file, and CIAO 4.8 \citep{Fruscione06}.
We reprocessed the observation using the \texttt{chandra\_repro} script and obtained a new level 2 events file.
In order to obtain an astrometry-corrected image, we compared the source list obtained from the long observation ObsID 9555 from \texttt{wavdetect} to \textit{Gaia} Data Release 2 and computed coordinate transforms using \texttt{wcs\_match} and \texttt{wcs\_update} \citep{GaiaDR2}.
We then corrected the astrometry of the other, shorter observations to match ObsID 9555 by the same procedure.
Exposure-corrected images were created for each observation using \texttt{fluximage}.
We merged the images using \texttt{merge\_obs}.
We also used \texttt{makepsf} to create individual PSF maps for each observation, and created an exposure-map weighted average PSF map for use in final source detection.
Sources were detected on the final astrometry-corrected, merged image, shown in Figure \ref{xray_des_image}, using \texttt{wavdetect}.
In \texttt{wavdetect} we detect wavelets at scales of 2, 4, 8, 16, 24, 32, and 48 pixels.

\begin{figure*}
    \centering
    \includegraphics[width=0.5\textwidth]{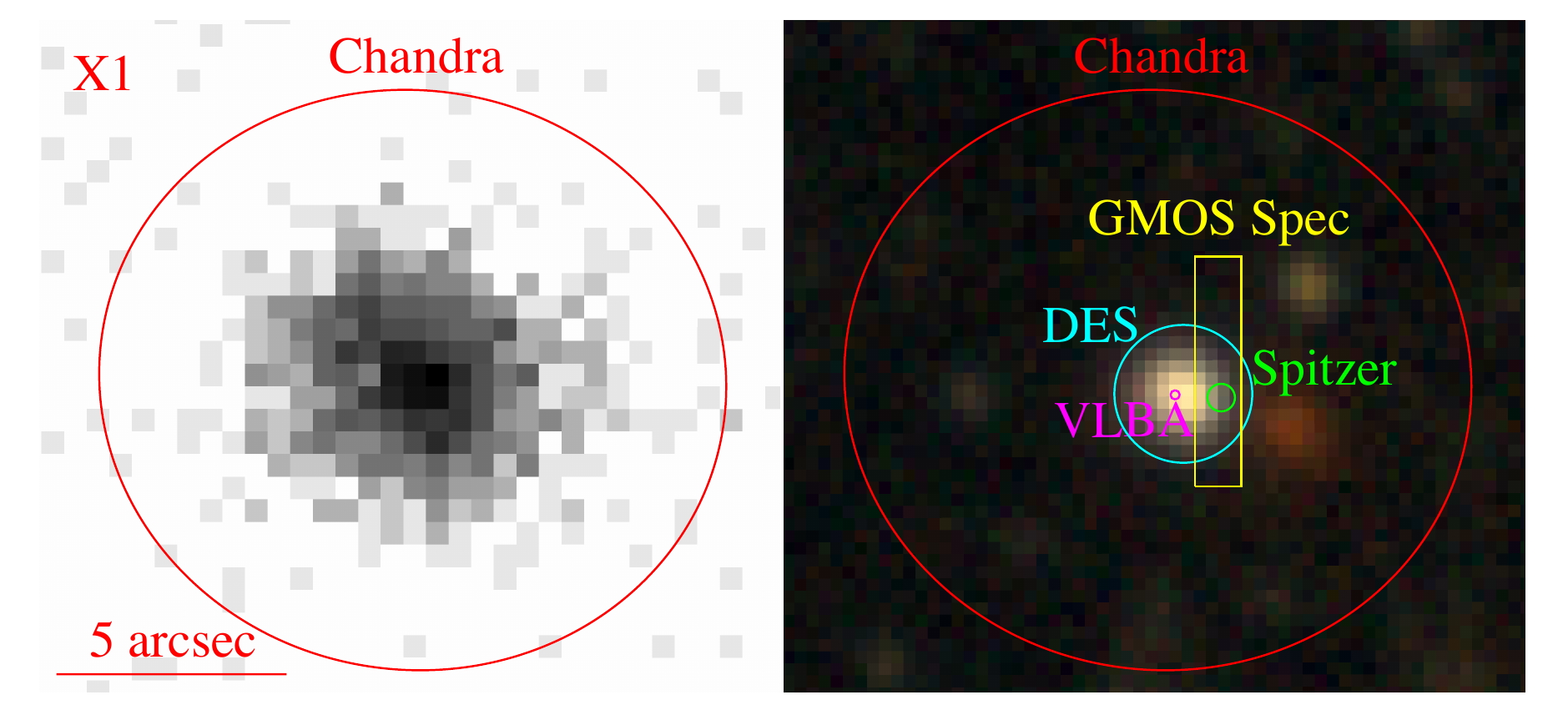}
    \caption{Left: \textit{Chandra} 0.5--10.0~keV image of SD X-1. Red circle marks the 3$\sigma$ source region for \textit{Chandra} determined by \texttt{wavdetect}. Right: Dark Energy Survey combined $g$ (blue), $r$ (green), and $i$ (red) image. Circles and rectangular regions mark positions of counterparts and their corresponding uncertainties. The DES image is set to the same scale as the corresponding \textit{Chandra} image.
    }
    \label{stamps1}
\end{figure*}

\begin{figure*}
    \centering
    \begin{subfigure}[t]{0.48\textwidth}
        \centering
        \includegraphics[width=0.95\linewidth]{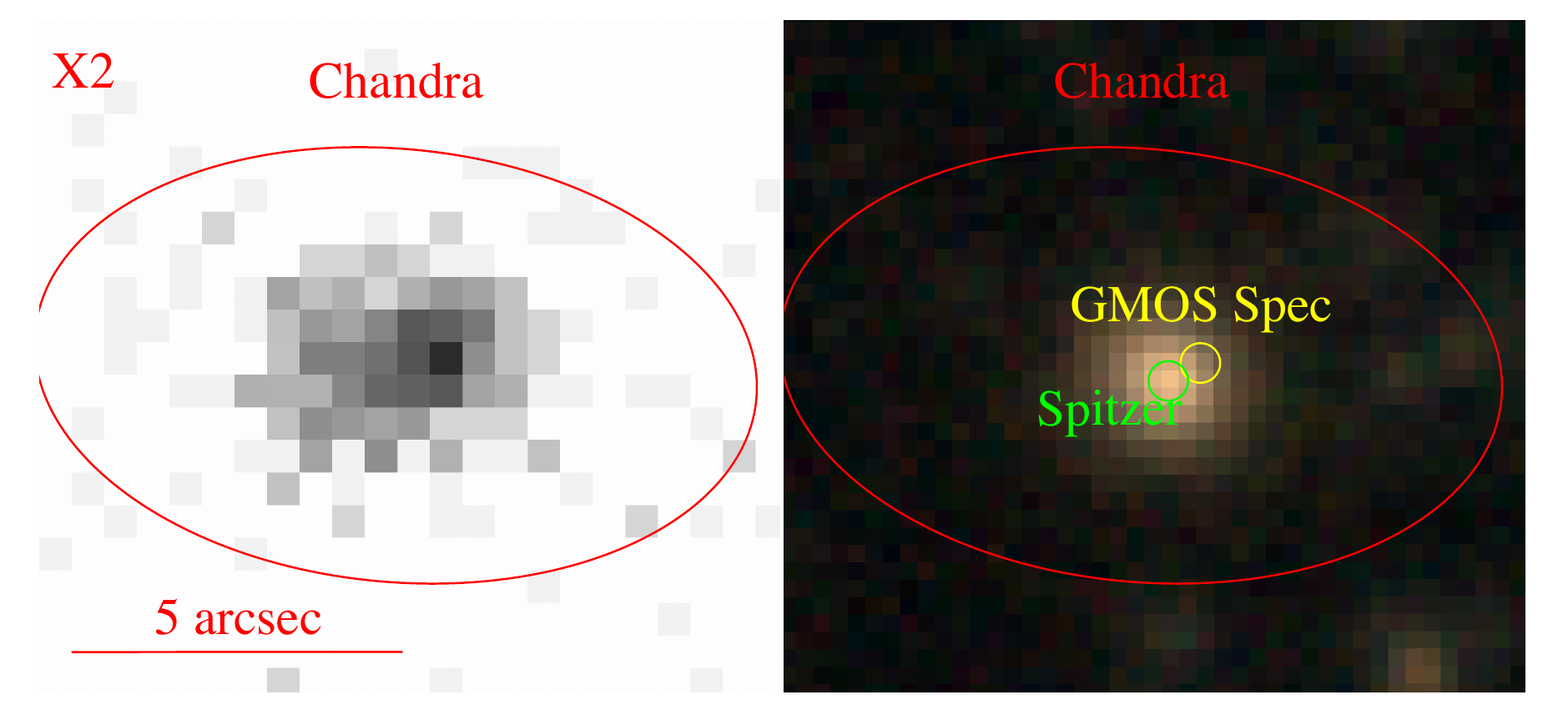}
        \caption{SD X-2} \label{x2stamp}
    \end{subfigure}
    \hfill
    \begin{subfigure}[t]{0.45\textwidth}
        \centering
        \includegraphics[width=0.95\linewidth]{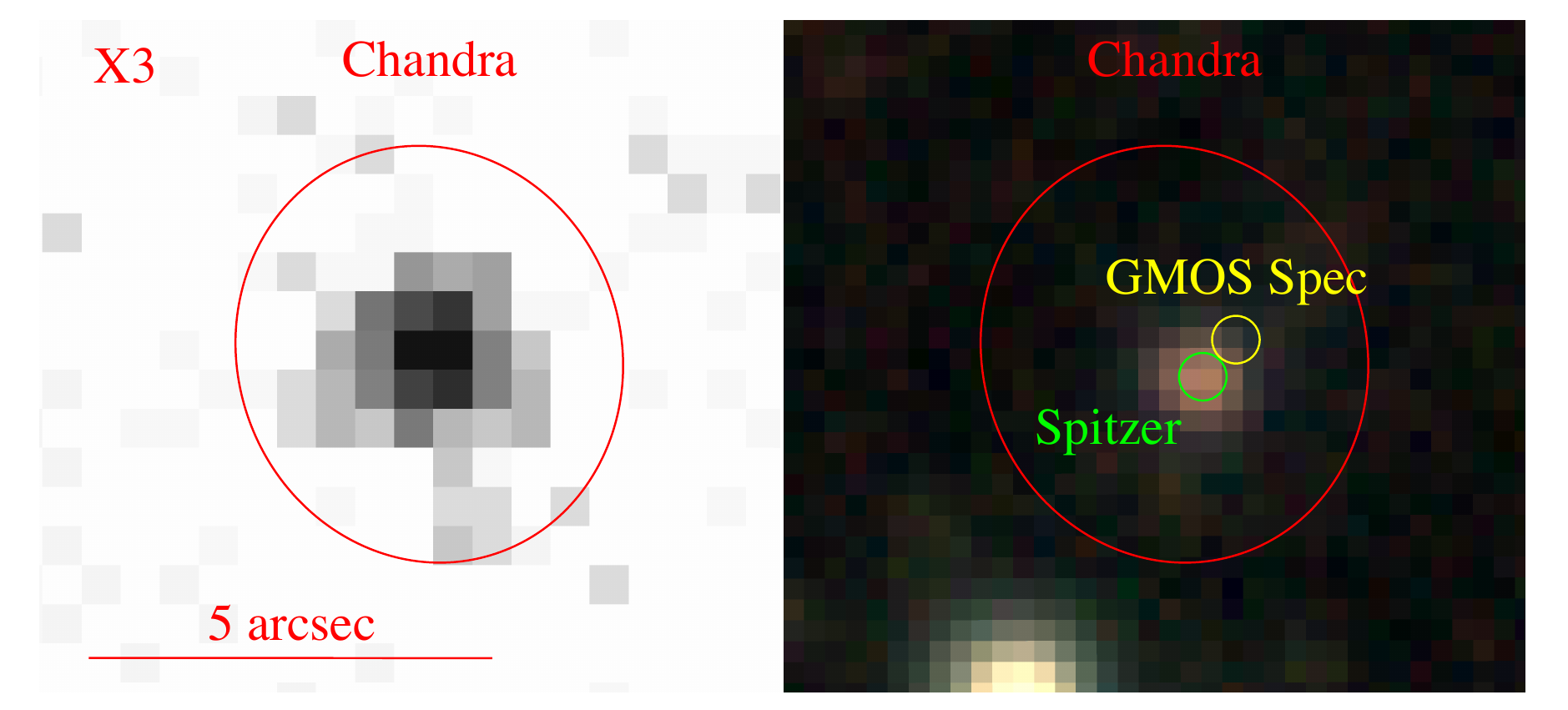}
        \caption{SD X-3} \label{x3stamp}
    \end{subfigure}

    \vspace{1cm}

    \begin{subfigure}[t]{0.48\textwidth}
        \centering
        \includegraphics[width=0.95\linewidth]{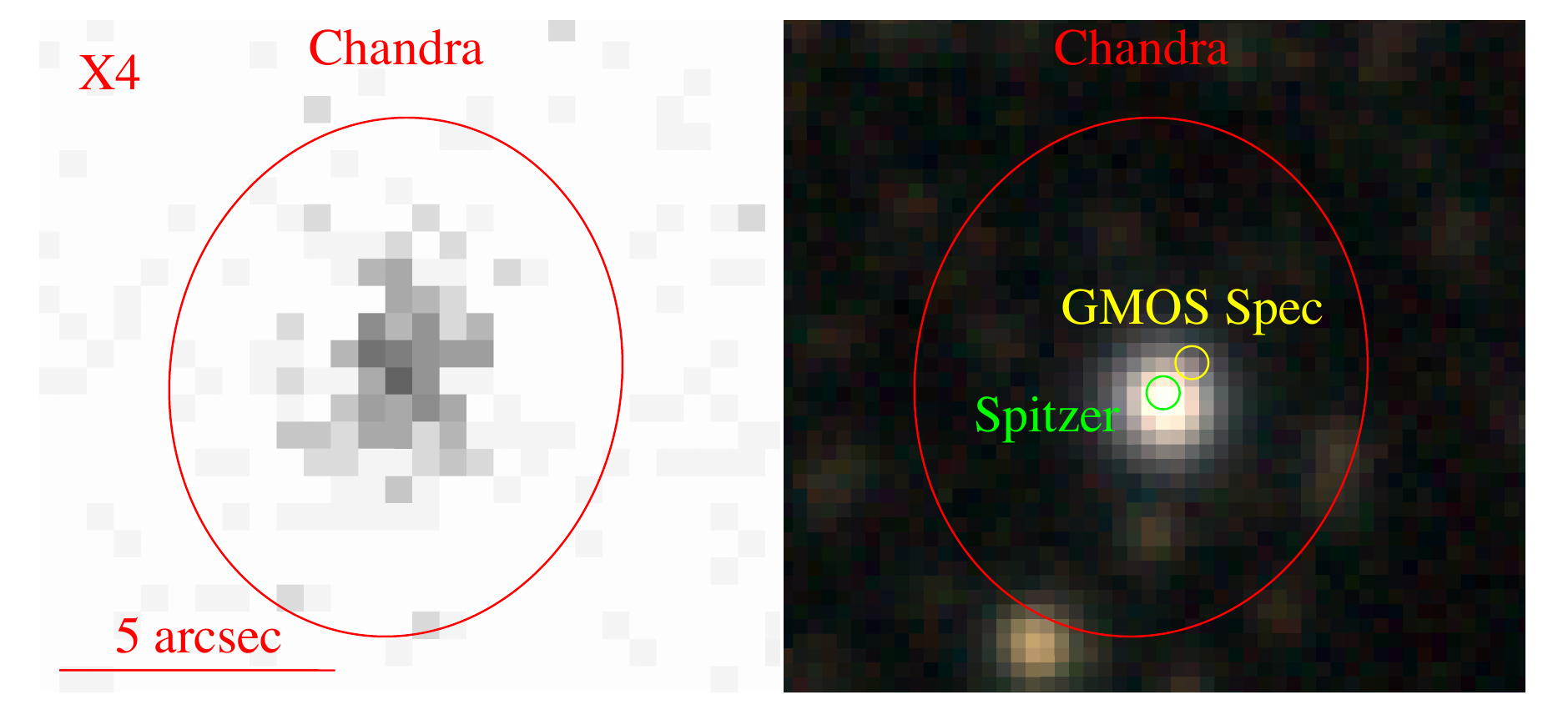}
        \caption{SD X-4} \label{x4stamp}
    \end{subfigure}
    \hfill
    \begin{subfigure}[t]{0.45\textwidth}
        \centering
        \includegraphics[width=0.95\linewidth]{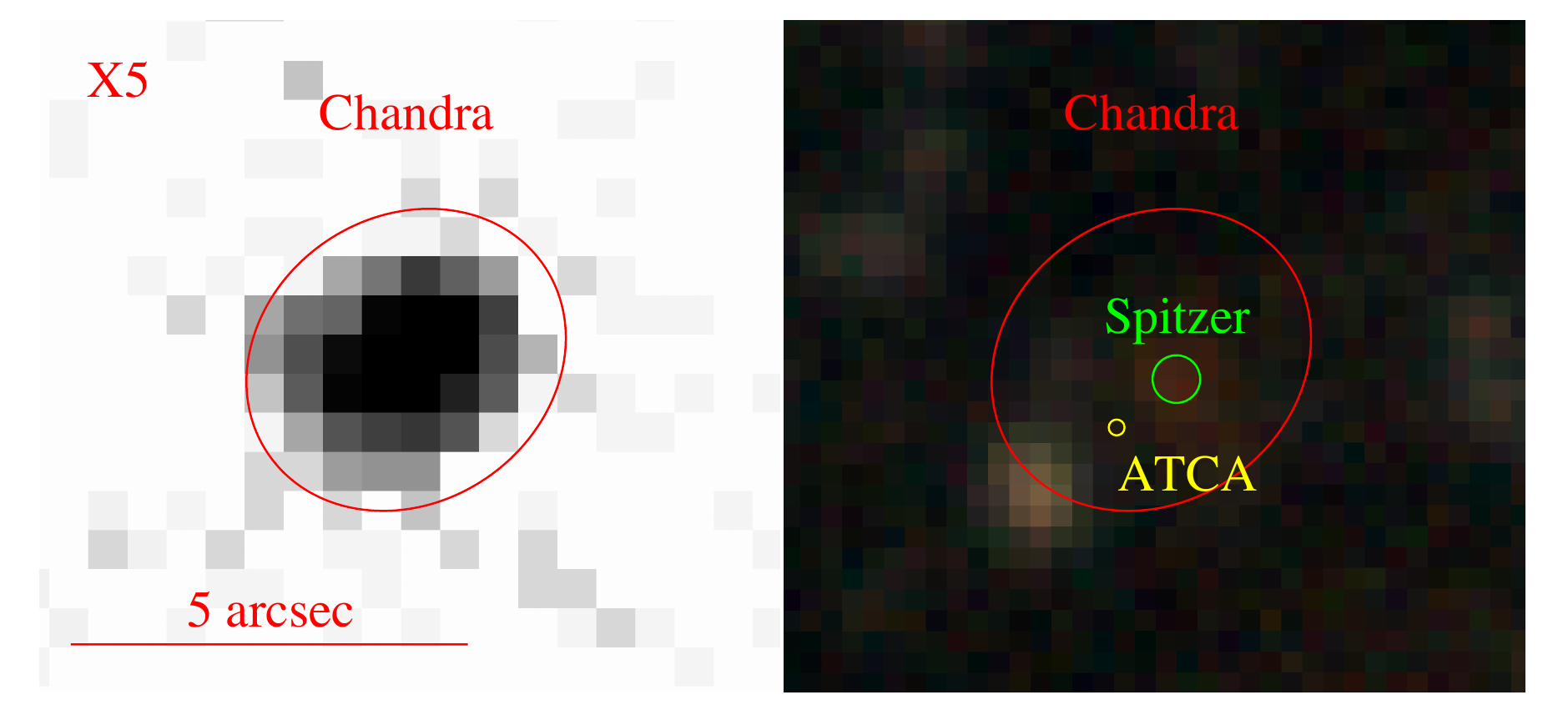}
        \caption{SD X-5} \label{x5stamp}
    \end{subfigure}

    \vspace{1cm}

    \begin{subfigure}[t]{0.48\textwidth}
        \centering
        \includegraphics[width=0.95\linewidth]{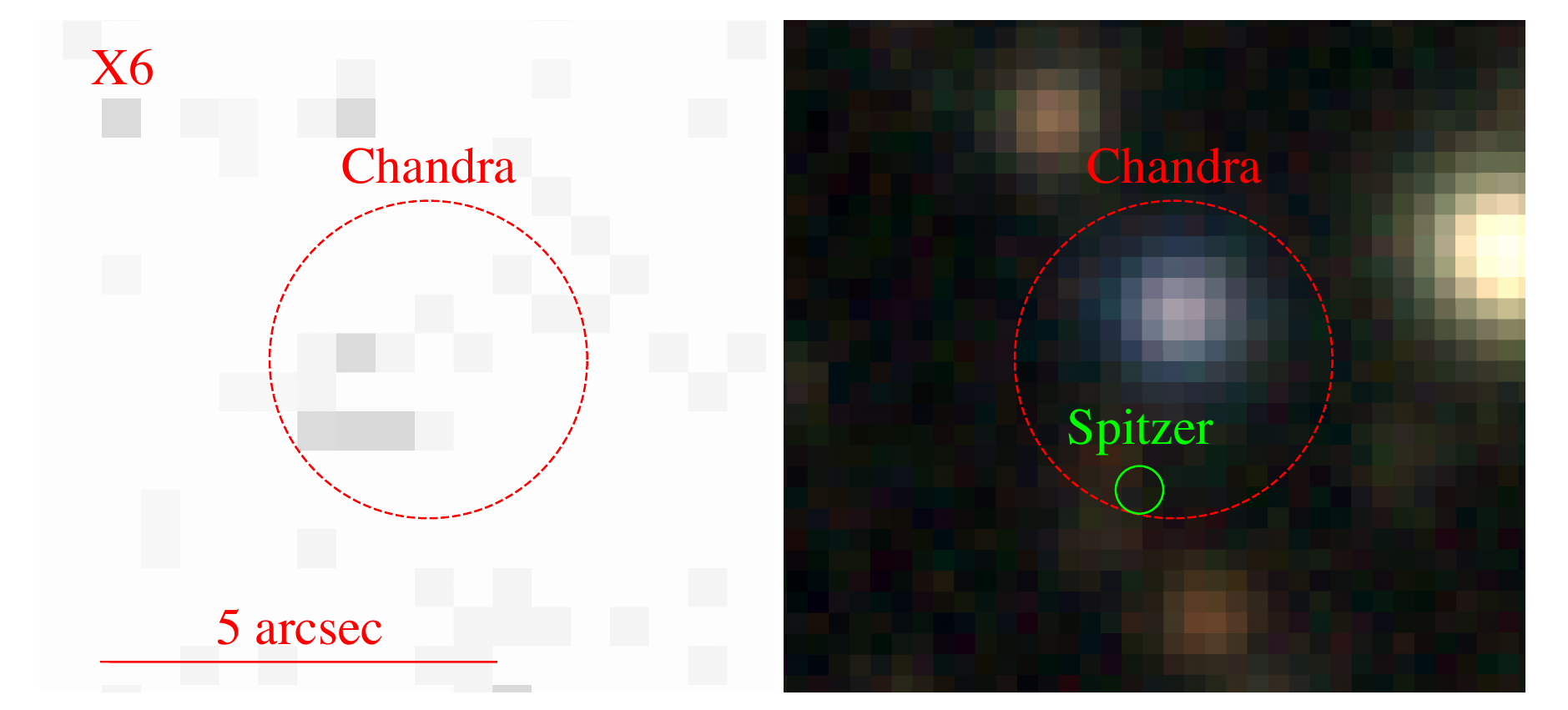}
        \caption{SD X-6} \label{x6stamp}
    \end{subfigure}
    \hfill
    \begin{subfigure}[t]{0.45\textwidth}
        \centering
        \includegraphics[width=0.95\linewidth]{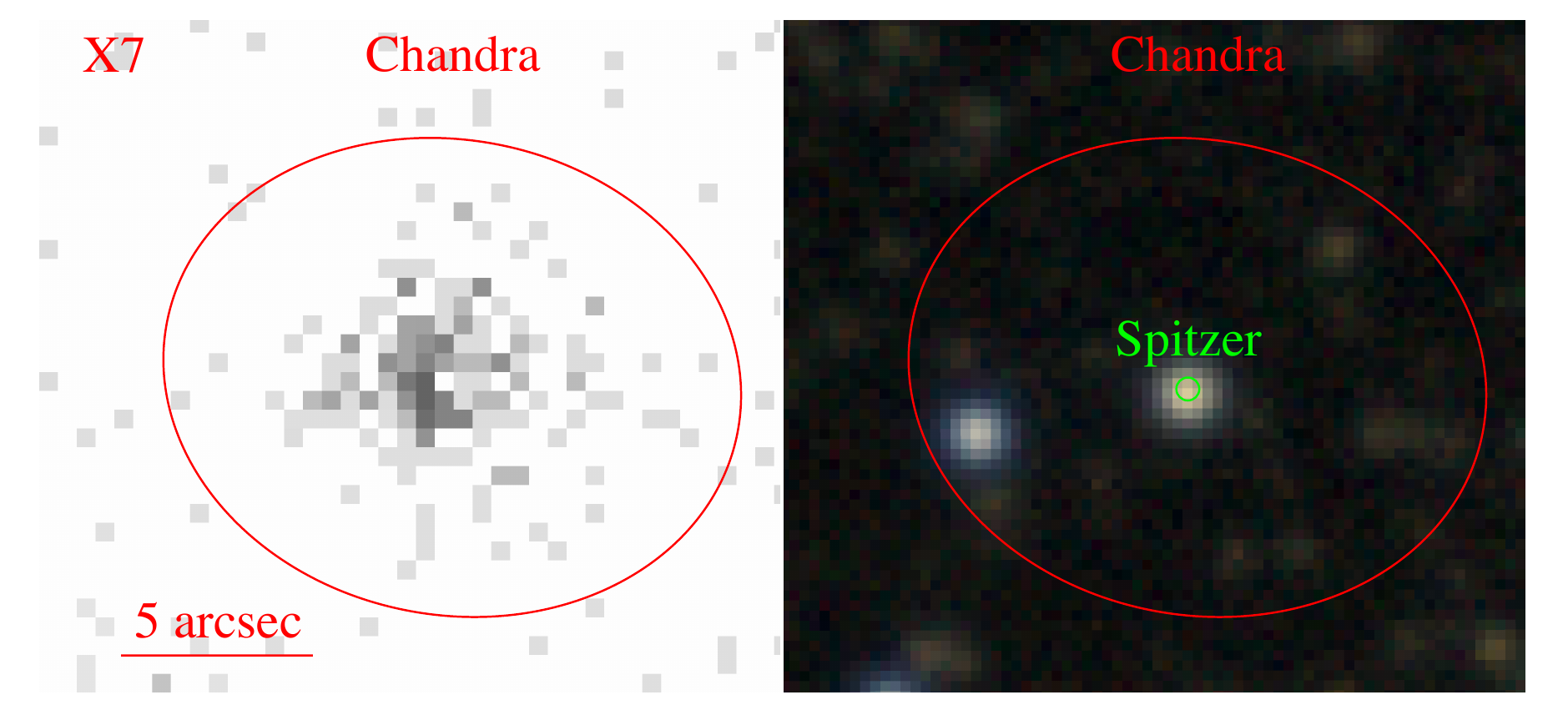}
        \caption{SD X-7} \label{x7stamp}
    \end{subfigure}
    \caption{\textit{Chandra} X-ray 0.5--10.0~keV images and Dark Energy survey combined $g$ (blue), $r$ (green), and $i$ (red) images of X-ray sources X-2 through X-7. Red circles mark the 3$\sigma$ X-ray source region as determined by \texttt{wavdetect}, while other circles mark the locations and corresponding uncertainties of multiwavelength counterparts. Each DES image is set to the same scale as the corresponding \textit{Chandra} image. }
    \label{stamps2}
\end{figure*}

\begin{figure*}
    \centering
    \begin{subfigure}[t]{0.48\textwidth}
        \centering
        \includegraphics[width=0.95\linewidth]{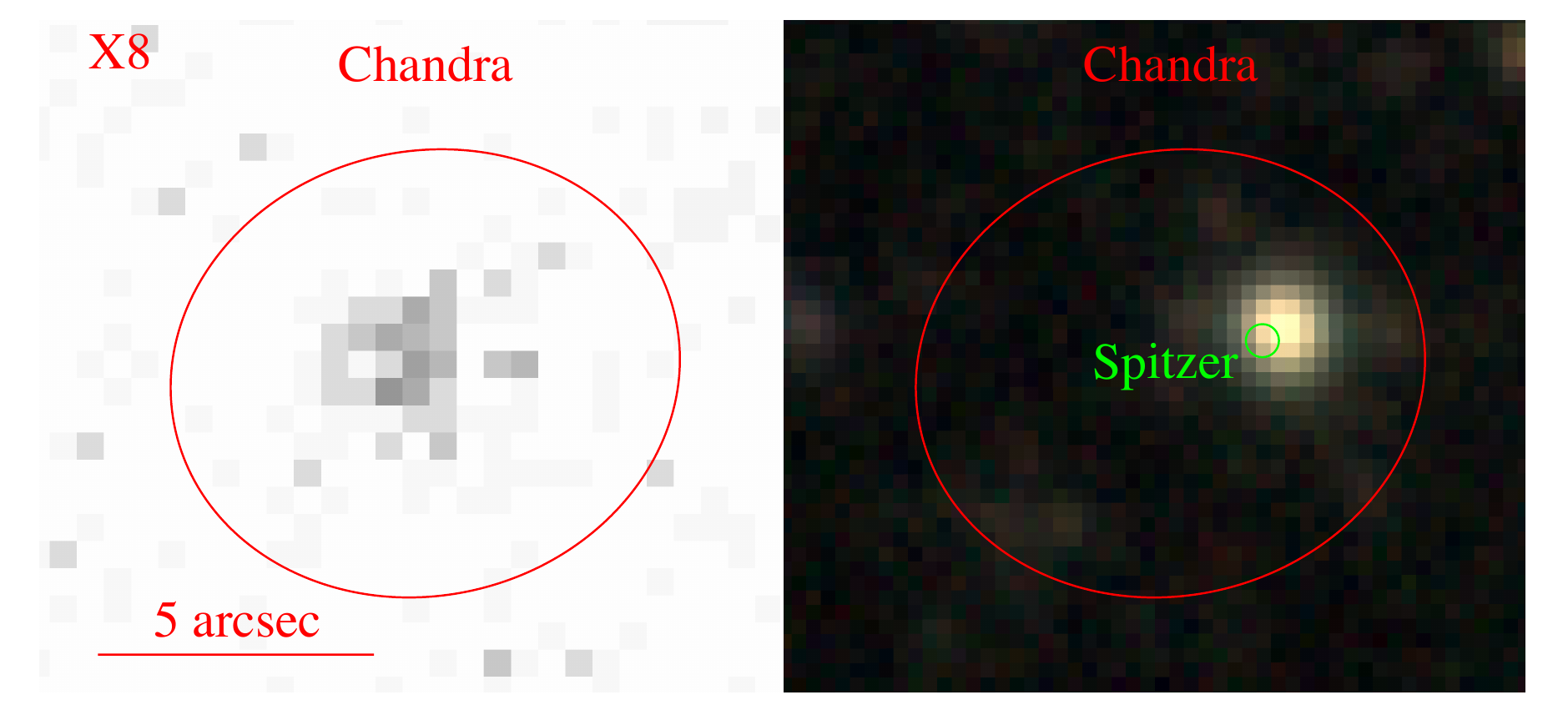}
        \caption{SD X-8} \label{x8stamp}
    \end{subfigure}
    \hfill
    \begin{subfigure}[t]{0.45\textwidth}
        \centering
        \includegraphics[width=0.95\linewidth]{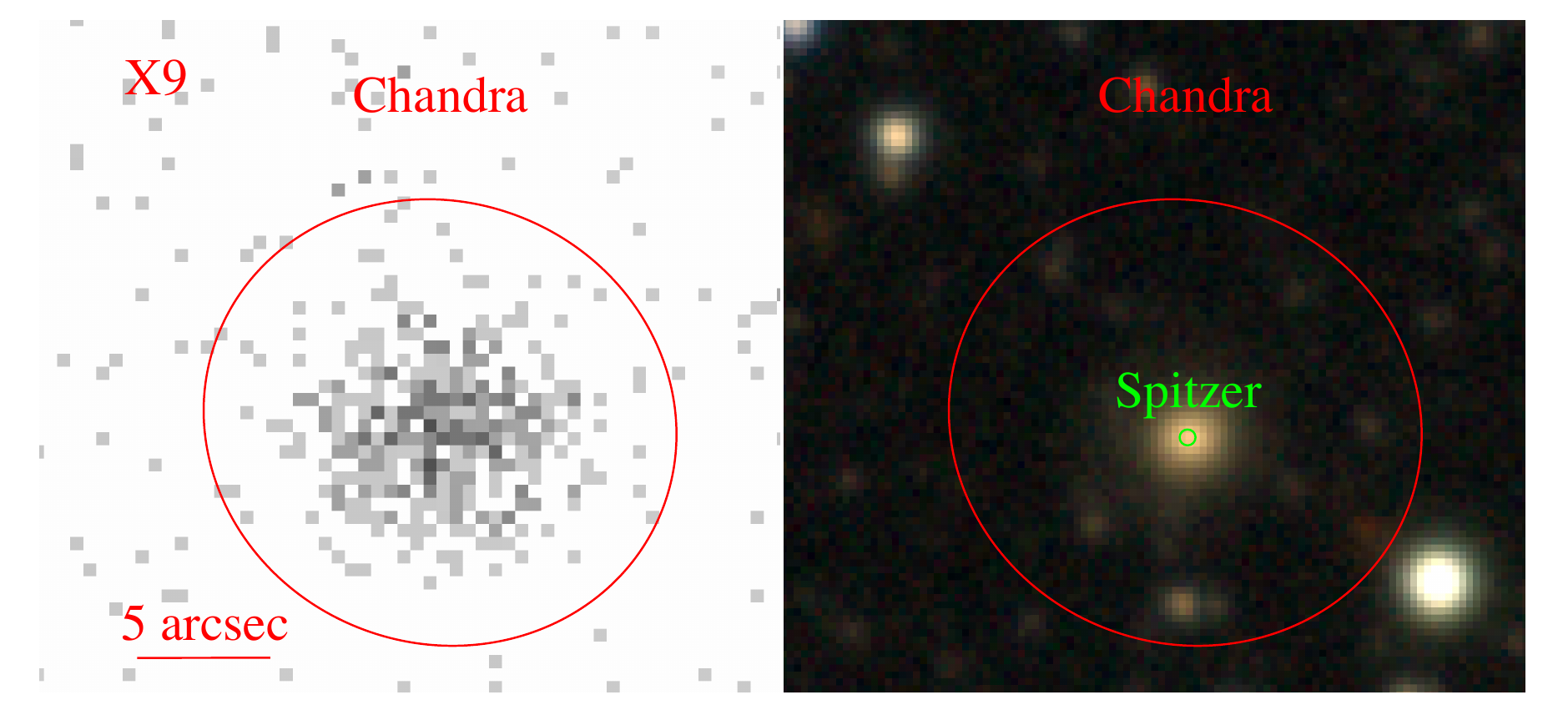}
        \caption{SD X-9} \label{x9stamp}
    \end{subfigure}

    \vspace{1cm}

    \begin{subfigure}[t]{0.48\textwidth}
        \centering
        \includegraphics[width=0.95\linewidth]{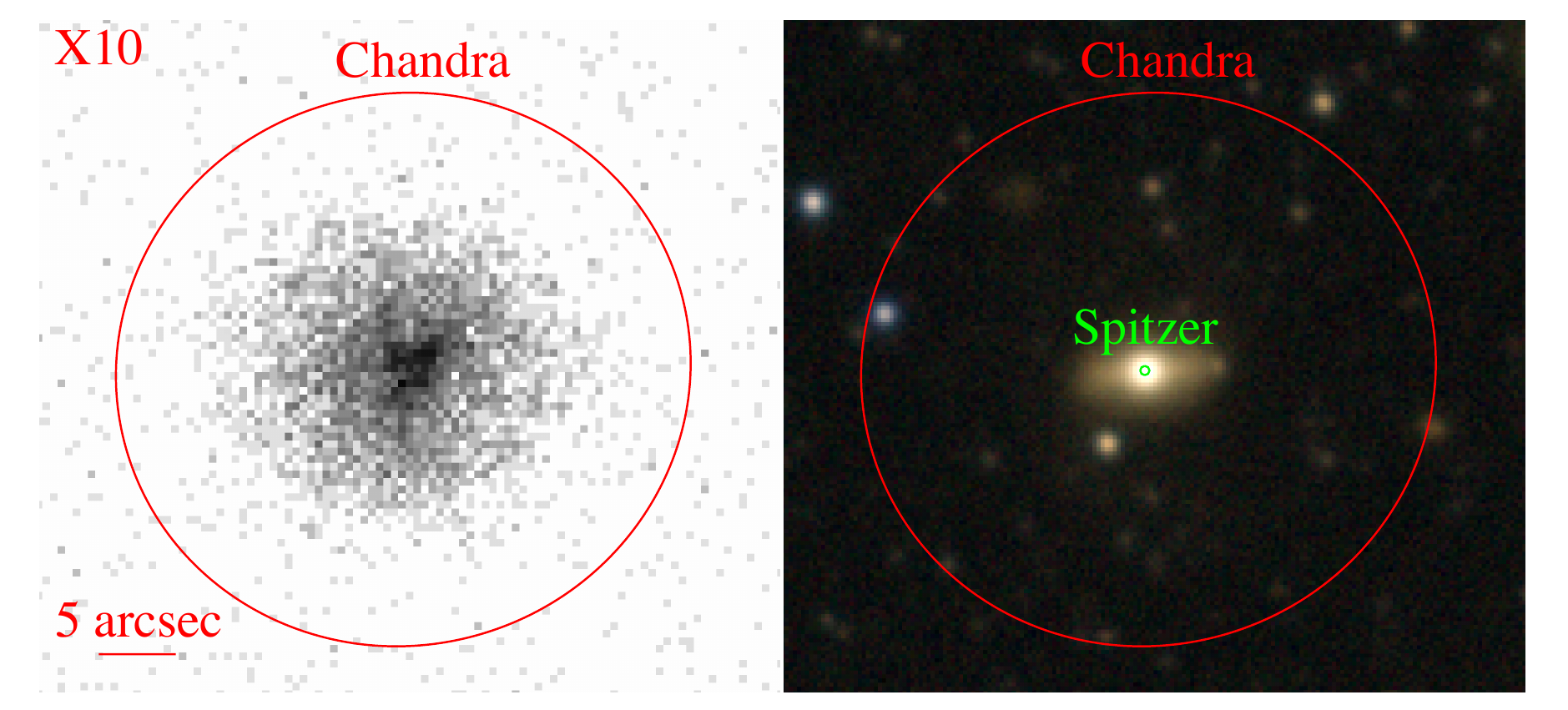}
        \caption{SD X-10} \label{x10stamp}
    \end{subfigure}
    \hfill
    \begin{subfigure}[t]{0.45\textwidth}
        \centering
        \includegraphics[width=0.95\linewidth]{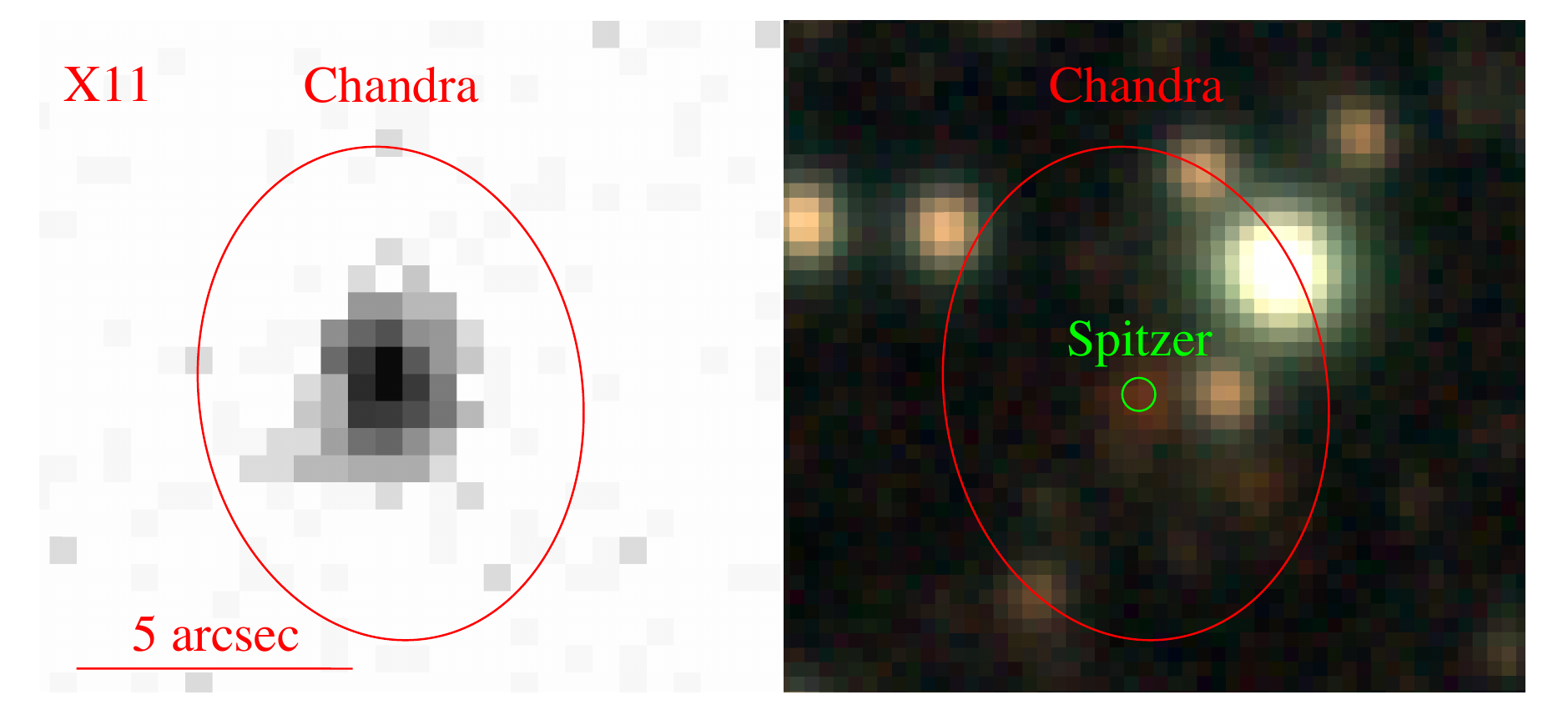}
        \caption{SD X-11} \label{x11stamp}
    \end{subfigure}

    \vspace{1cm}

    \begin{subfigure}[t]{0.48\textwidth}
        \centering
        \includegraphics[width=0.95\linewidth]{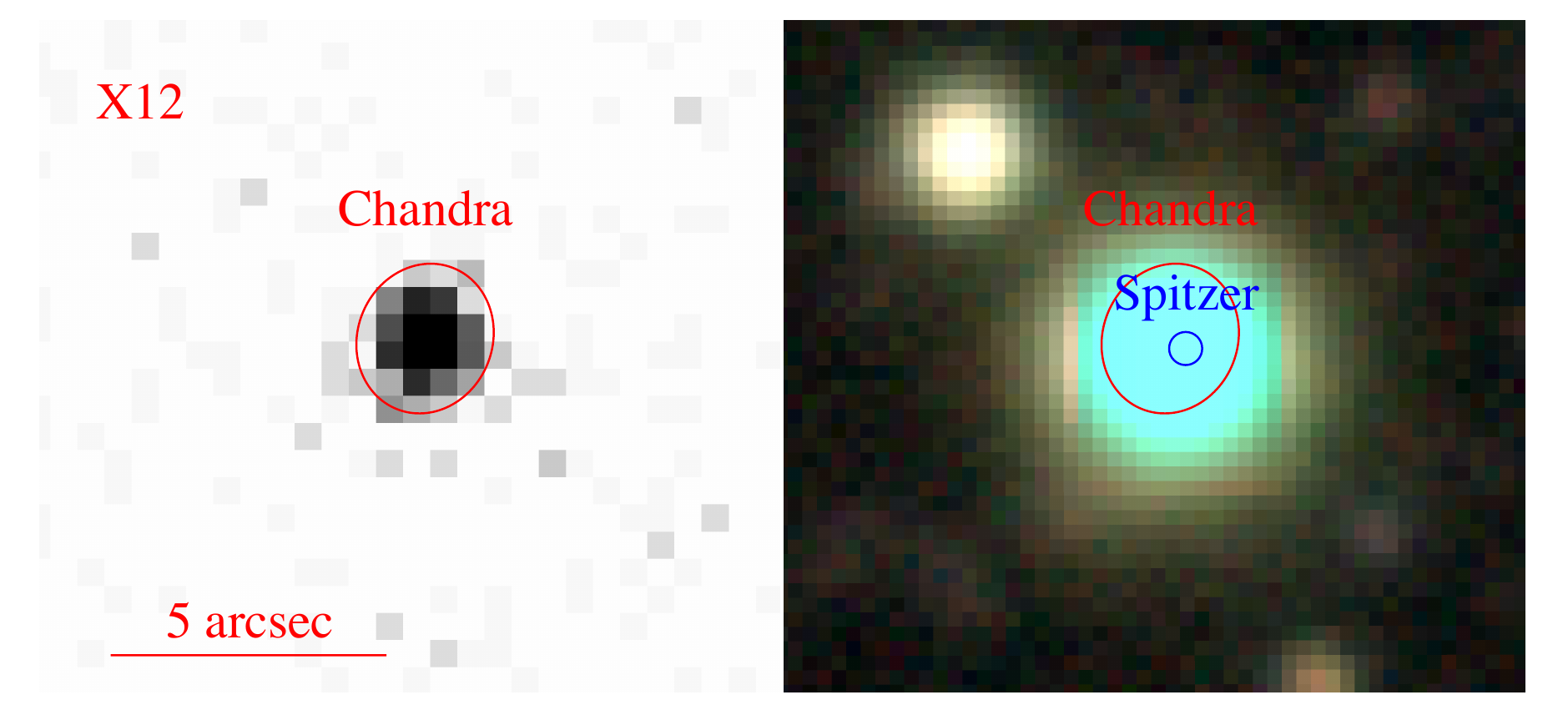}
        \caption{SD X-12} \label{x12stamp}
    \end{subfigure}
    \hfill
    \begin{subfigure}[t]{0.45\textwidth}
        \centering
        \includegraphics[width=0.95\linewidth]{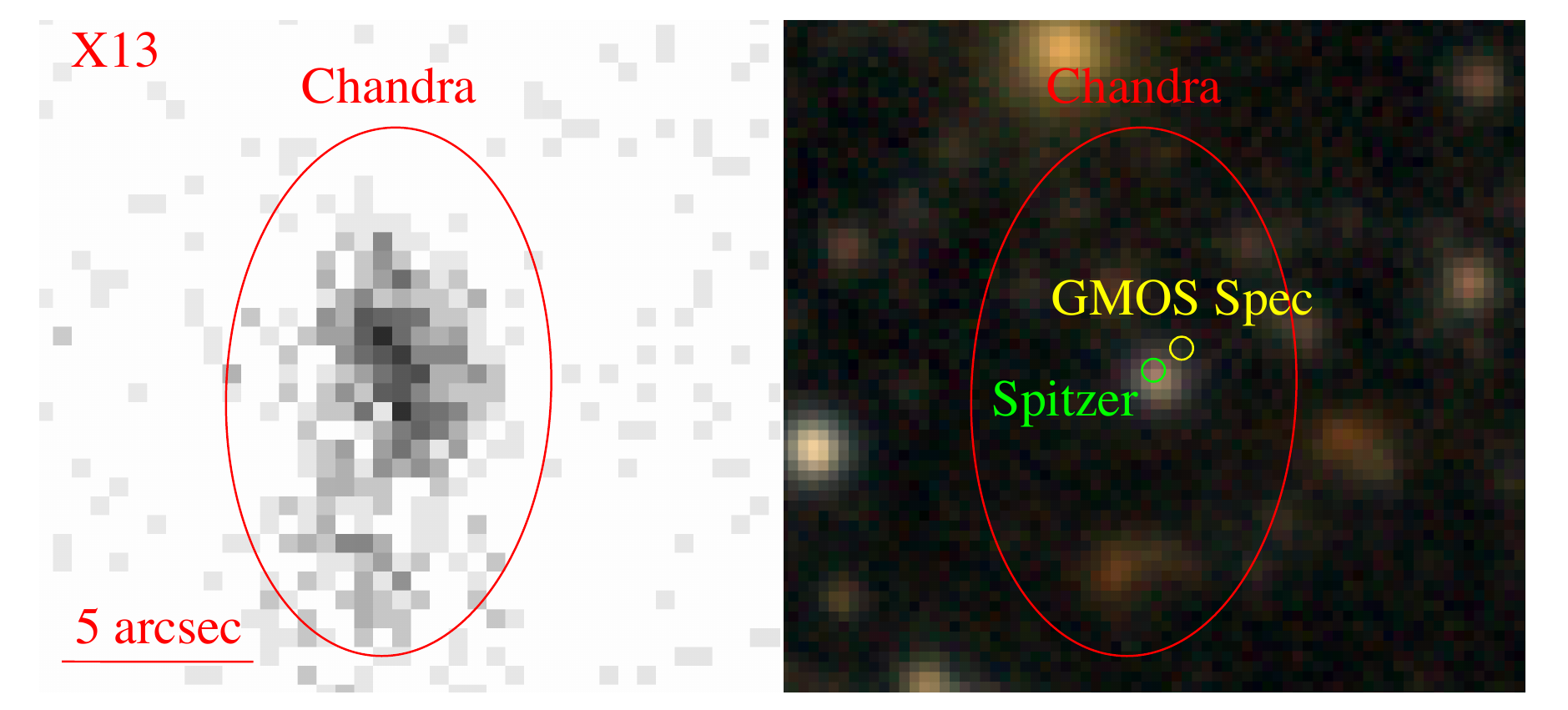}
        \caption{SD X-13} \label{x13stamp}
    \end{subfigure}

    \caption{\textit{Chandra} X-ray 0.5--10.0~keV images and Dark Energy survey combined $g$ (blue), $r$ (green), and $i$ (red) images of X-ray sources X-8 through X-13. Red circles mark the 3$\sigma$ X-ray source region as determined by \texttt{wavdetect}, while other circles mark the locations and corresponding uncertainties of multiwavelength counterparts. Each DES image is set to the same scale as the corresponding \textit{Chandra} image.
    }
    \label{stamps3}
\end{figure*}

We considered two groups of sources in this analysis for individual study.
Firstly, we considered the sources previously detected in \cite{Maccarone05a}.
We also investigated any source with $>100$ source counts in the long observation, as these sources would be candidates for spectral analysis in the event that they appeared to be promising XRB candidates.
Applying \texttt{wavdetect} to the combined image detects eight of the nine sources from the \cite{Maccarone05a} study (all except X6), and detects four new sources that meet the $>100$ source count criterion.
The positions, previous identifications, and X-ray properties of these 13 sources are summarized in Tables \ref{sourcetablepos} and \ref{sourcetablexray}.
X-ray images of the sources are shown in Figures~\ref{stamps1}, \ref{stamps2}, and \ref{stamps3}.
The naming scheme used is such that SD X-1 through SD X-9 are the \cite{Maccarone05a} sources, listed in the same order as in that paper's tables 1 and 2.
Sources SD X-10 to SD X-13 are the newly-detected bright sources.
Matching the source positions to catalogues at other wavelengths is discussed below, in section \ref{matching}.

We also use the \texttt{wavdetect} source list to construct a combined X-ray luminosity function for the Sculptor field of view. In order to convert the \texttt{net\_rate} count rate measured by \texttt{wavdetect} into an unabsorbed flux, we assume a power law of index $\Gamma = 2.0$ and column density of $N_{\rm H} = 9.0 \times 10^{19}$~cm$^{−2}$.
The resulting X-ray source counts are plotted in Figure \ref{lognlogs}.
For the bright sources of interest, we crossmatched these sources to the \textit{Chandra Source Catalog}, using TOPCAT \citep{TOPCAT} and a 2\arcsec matching tolerance.
Agreement between the two source lists was good, with a mean astrometric offset of 0.8\arcsec.
The above analysis using the combined \textit{Chandra} observations is useful for constraining source astrometry and morphology, though we defer to the \textit{Chandra Source Catalog} to obtain fluxes and hardness ratios for bright individual sources of interest, shown in Table \ref{sourcetablexray}.

\begin{table*}
\caption{%
Summary of \Chandra\ X-ray observations of the Sculptor Dwarf Spheroidal Galaxy. All observations were made using ACIS-S.
}
\label{xrayobs}
\begin{tabular}{lccccc}
\hline \hline
ObsID & Date & Exposure Time & PI & Mode \\
 & & ks & & \\
\hline
4698 & 2004-04-26 08:33:34 & 6.06 & Maccarone & VFAINT \\
4699 & 2004-05-07 04:39:34 & 6.27 & Maccarone & VFAINT \\
4700 & 2004-05-17 10:26:45 & 6.1 & Maccarone & VFAINT \\
4701 & 2004-05-30 22:16:43 & 6.07 & Maccarone & VFAINT \\
4702 & 2004-06-12 14:19:32 & 5.88 & Maccarone & VFAINT \\
4703 & 2004-06-27 20:08:18 & 5.88 & Maccarone & VFAINT \\
4704 & 2004-07-12 01:08:02 & 5.91 & Maccarone & VFAINT \\
4705 & 2004-07-24 07:51:23 & 5.83 & Maccarone & VFAINT \\
4706 & 2004-08-04 11:08:39 & 6.08 & Maccarone & VFAINT \\
4707 & 2004-08-17 04:52:59 & 5.88 & Maccarone & VFAINT \\
4708 & 2004-08-31 05:46:02 & 5.88 & Maccarone & VFAINT \\
4709 & 2004-09-16 03:47:43 & 6.09 & Maccarone & VFAINT \\
4710 & 2004-10-01 10:50:52 & 5.88 & Maccarone & VFAINT \\
4711 & 2004-10-11 14:32:47 & 5.88 & Maccarone & VFAINT \\
4712 & 2004-10-24 01:32:41 & 6.08 & Maccarone & VFAINT \\
4713 & 2004-11-05 02:28:24 & 6.07 & Maccarone & VFAINT \\
4714 & 2004-11-20 18:48:32 & 6.04 & Maccarone & VFAINT \\
4715 & 2004-12-05 06:40:06 & 5.68 & Maccarone & VFAINT \\
4716 & 2004-12-19 08:46:48 & 6.01 & Maccarone & VFAINT \\
4717 & 2004-12-29 23:47:24 & 6.07 & Maccarone & VFAINT \\
4718 & 2005-01-10 04:08:34 & 6.06 & Maccarone & VFAINT \\
9555 & 2008-09-12 00:11:10 & 49.53 & Zepf & FAINT \\
\end{tabular}
\end{table*}

\subsection{Gemini Imaging Data Reduction}
We obtained observations of the core of Sculptor using Gemini GMOS-S in the r and H$\alpha$ bands (PID: GS-2016B-Q66, Dates: 2016-08-14 \& 2016-08-15, PI: R. Arnason).
The central $5.5 \times 5.5$~arcmin of Sculptor was imaged using $20 \times 500$~s H$\alpha$ exposures and $8 \times 180$~s $r$ exposures.
The observations were split between two nights, only one of which had sufficiently good seeing to be usable.
Therefore, our final images are comprised of $10 \times 500$~s in H$\alpha$ and $4 \times 180$~s in $r$.
From use of the IRAF task \texttt{imexamine}, we estimate the FWHM of these images to be roughly 4 pixels, combined with 0.16\arcsec per pixel gives a FWHM of 0.64\arcsec.
The reduction was performed using PyRAF v2.2dev.
We derived normalized flatfields and bias frames using GIFLAT and GBIAS.
All images were reduced using GIREDUCE, and we corrected a defect
on the GMOS-S detector using an observatory-supplied script (G. Gimeno, private communication).
The images were then mosaiced using GMOSAIC and coadded using IMCOADD to create the final images, which had $2\times2$ binning and a 0.16$\arcsec$ pixel size.

To detect and extract sources, we used SExtractor 2.19.5.
Source extraction was done with the following settings: 5$\arcsec$ fixed aperture, a $3\sigma$ detection threshold, and a minimum of 8 pixels above the threshold required for source detection.
We derived an $r$-band magnitude calibration through comparison with $r$-band catalogues in the Dark Energy Survey (DES).
We cross-matched sources in our imaging with those in the DES images, and derived a constant magnitude correction that we apply to match SExtractor's magnitudes to those measured by DES.
For the H$\alpha$ data, we fit a line through the main sequence of the $r$ vs. H$\alpha$ - $r$ colour-magnitude diagram.
We then calculated a constant magnitude correction such that the main sequence had ${\rm H}\alpha - r = 0$.
As such, our H$\alpha$ magnitudes are relative and not an absolute calibration.
We calculate the photometric uncertainties using the SExtractor defaults.
The resulting catalogue was matched to the X-ray catalogue with TOPCAT using a 1$\arcsec$ tolerance.

\subsection{Gemini Spectroscopic Data}
We have obtained observations of two fields in Sculptor with GMOS-S in Multi-object spectroscopy (MOS) mode (PID: GS-2008B-Q-25, Date: 2008-09-21 PI: S. Zepf). The spectra were taken using the B600 G5323 grating, with 3 exposures of 420~s per field. The spectra have a resolution of approximately 8 \r{A} (as measured from the FWHM of the arc lamp lines) and a wavelength range of roughly 4000--7000 \r{A}. Four of the sources discovered by \cite{Maccarone05a} were observed in this program. Each of the three exposures were taken with slightly different central wavelengths in order to remove the GMOS chip gaps from the final spectra. We derived normalized spectral flats and bias frames with GSFLAT and GBIAS. Arc, standard, and science images were created using GSREDUCE. Each exposure was filtered for cosmic rays using GEMCRSPEC. The images were wavelength calibrated using GSWAVELENGTH, and sky subtracted using GSSKYSUB. We extracted 1D spectra from the 2D spectra using GSEXTRACT, and then stacked the exposures together with GEMSCOMBINE. The stacked exposures were then flux calibrated using   GSSTANDARD and the spectrum of the standard star (LTT1020). The resulting final spectra were analyzed and fit using SHERPA 4.9.0.

\subsection{\Spitzer\ Data}
Mid-infrared images of Sculptor were obtained in \Spitzer\ Cycle 5 (PID: 50314, PI: P. Barmby).
Infrared Array Camera \citep[IRAC;][]{Fazio2004} observations were made in 2008 December and covered the galaxy at wavelengths of 3.6, 4.5, 5.8, and 8.0 $\mu$m in a $0.4\times 0.7$~deg map with $5\times12$~s dithered observations per sky position.
The expected $5\sigma$ point source detection limits are roughly 10~$\mu$Jy at 3.6 and 4.5~$\mu$m and 100~$\mu$Jy at 5.8 and 8.0 $\mu$m.
The IRAC point-spread function FWHM is approximate 1.8--2.0\arcsec.
Multiband Imager and Photometer for Spitzer \citep[MIPS;][]{Rieke2004} observations were made in 2008 August in scan-map mode, using medium scan speed with half-array offsets to cover an area of $0.4\times 1.2$~deg.
While data were collected in all three of the MIPS bands, here we focus on only the 24~$\mu$m data as the spatial resolution of the 70 and 160~$\mu$m bands is low compared to the other wavelengths of interest.
The expected $5\sigma$ point source detection limit is roughly 800~$\mu$Jy at 24~$\mu$m, with a PSF FWHM of 6\arcsec.

A detailed description of data processing and full catalogues are in a forthcoming paper (Barmby et al. 2019, in preparation); here we give a brief summary.
IRAC images were processed using pipeline S17.0.1.
The Level 1 basic calibrated data (BCD) files were cleaned with custom cleaning scripts to remove artifacts.
Images were mosaiced using \textsc{IRACproc} post-BCD Processing package 4.0, removing transient events and fixed-pattern background noise \citep{Schuster06}.
Source extraction of IRAC mosaics was performed using SExtractor and the following settings: detection threshold of $1.5\sigma$, 5 pixel minimum area, apertures of 2.46$\arcsec$, 3.66$\arcsec$, and 6.08$\arcsec$ with aperture corrections from the IRAC Instrument Handbook \citep{IRACHandbook}.
Aperture photometry was done with IRAF/apphot, and photometric uncertainties were estimated using apphot's standard formula.
MIPS images were processed using the MIPS Data Reduction tool.
Point sources were extracted for all MIPS bands using the PSF-fitting program StarFinder with model PSFs from \cite{Engelbracht07a}.
MIPS sources were matched to IRAC 3.6 $\mu$m sources using TOPCAT.
Most  ($\sim 80$\%) MIPS sources within  the IRAC coverage area had IRAC counterparts (those without were most often located near bright stars), but as expected for normal stars, most IRAC sources do not have MIPS
counterparts.

\subsection{Matching}
\label{matching}
In order to classify sources detected in our X-ray catalogue, we have carefully considered cross-matching between the \Chandra\ detections, our own catalogues, and external catalogues retrieved from ViZieR.
We have also considered cross-matching with photometry from the Dark Energy Survey
\citep[DES;][]{Abbott18a}, which provides coverage of Sculptor in \textit{griz}Y filters.
We show the X-ray sources considered over-plotted on a DES image in Figure~\ref{xray_des_image}.
\Chandra's absolute positional uncertainty is $\sim0.8$\arcsec.
For ordinary on-axis point sources, we expect that associations should match within this tolerance.
However, we chose a larger tolerance of $2\arcsec$ to account for several factors within the dataset.
First, a number of the sources in the field lie off-axis, resulting in greater positional uncertainty.
Second, at least one of the X-ray sources appears to be extended (see the section on SD X-10, below), and its optical counterpart is extended over several arcseconds.
Third, a number of the X-ray sources from \cite{Maccarone05a} are relatively faint or are detected only as upper limits, and as such, the positional accuracy is reduced.
We expect that point sources in our catalogue should match within $\sim0.8$\arcsec, and permit a higher tolerance only for these extended, faint, or off-axis sources.
As delivered, the Gemini GMOS $r$ and H$\alpha$ images were misaligned in World Coordinate System (WCS) space.
Therefore, the astrometry was aligned by matching bright sources in the individual filter images to bright sources in the \Spitzer\ 3.6 $\mu$m images with TOPCAT and deriving an average shift.
We selected the 100 brightest sources in each image to derive the shift.
This correction only accounts for misalignments due to translation, and not rotation or scale corrections; we used the standard deviation of the difference between the \Spitzer\ and $r$ and H$\alpha$  images to estimate the size of the positional uncertainty due to rotation or scale.
For both the $r$ and H$\alpha$  images, we find that this uncertainty is roughly $\sim0.2\arcsec$.

\section{Analysis}

\begin{figure}
\includegraphics[width=0.45\textwidth]{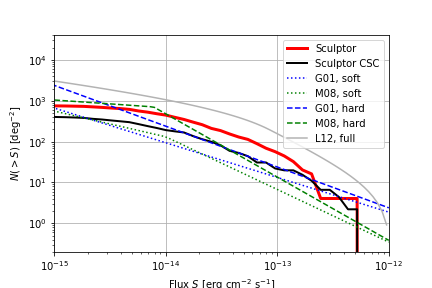}
\caption{%
Cumulative source counts derived from the Sculptor \Chandra\ observations and from fits to deep field observations by \protect\cite{Lehmer12}, \protect\cite{Giacconi01} and \protect\cite{Mateos05}.
Note that the Sculptor counts (from the combined observation and the \Chandra\ Source Catalog) and those from \protect\cite{Lehmer12} are in the full 0.5--8~keV band, while the  counts from \protect\cite{Giacconi01} and \protect\cite{Mateos05} are shown separately for the hard and soft bands.}
\label{lognlogs}
\end{figure}

Figure~\ref{lognlogs} presents the cumulative source counts derived from the \Chandra\ observations, together with comparison counts from deep-field observations from  \cite{Giacconi01}, \cite{Mateos05} and \cite{Lehmer12}.
The Sculptor field does not contain an excess of sources over the background counts, suggesting that there are few if any bright X-ray sources within the galaxy.
The small number of sources means that we can go beyond the statistical argument to examine each bright source individually to investigate its association with Sculptor.

In our new observation, We re-detect six of the nine bright sources from \cite{Maccarone05a}, and we identify four new sources that have $>100$ source counts.
Where possible, we use the position obtained in our new observation with WAVDETECT.
For X-6, which was not detected in the new observation, we use the position from \cite{Maccarone05a}.
Note that we find the position of one source, SD X-8, to be different from its reported location in \cite{Maccarone05a} (see Section~\ref{sec:sdx8}, below).
To classify sources, we use X-ray and optical spectroscopy, IR and optical photometery, and comparison with catalogues.
For sources with \Spitzer\ photometry, we use the AGN selection wedge of \cite{Stern05a}, to identify sources as AGN.
Amongst the X-ray sources in the Sculptor dwarf field, we find that two are background galaxies, seven are background AGN, and the nature of three is uncertain.
In this section we discuss each of the X--ray sources individually, with the population properties explored in the following section.
A summary of literature IDs, classifications from the literature, and the classification derived in this work for each source is given in Table~\ref{sourcetablepos}.
Each source's X-ray properties are summarized in Table~\ref{sourcetablexray}, and the properties of its optical/mid-IR counterparts, if any, are given in Table~\ref{sourcetableiropt}.

\subsection{SD X-1}
The X-ray source at this position has existing counterparts in the literature.
Previous quasi-stellar object (QSO) catalogues have identified both a radio \citep{Tinney97a} and optical counterpart \citep{Perlman98a}. Previous optical spectroscopy by \cite{Perlman98a} identified a single broad optical feature at roughly 5250~\r{A} and a velocity width of roughly 3800~km~s$^{-1}$.
The relative isolation of the line led to its identification as \ion{Mg}{ii}.
This line, plus the observed radio flux, led to the subsequent identification of this source as a BL Lac with redshift $z=0.875$.

\begin{figure*}
\includegraphics[width=0.45\textwidth]{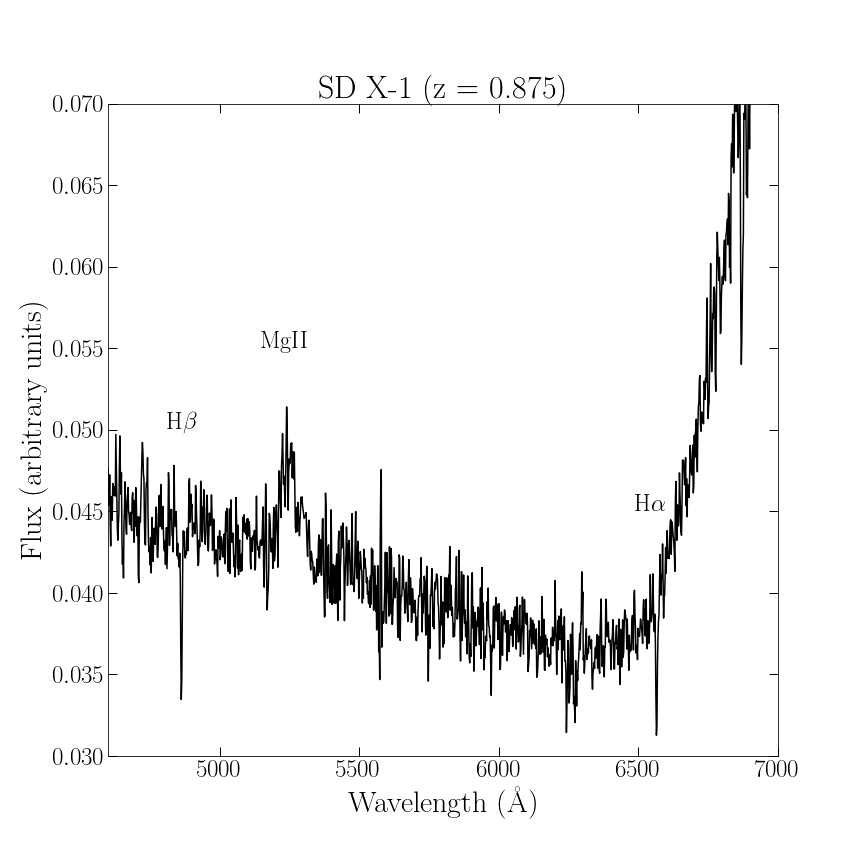}
\caption{%
GMOS-S MOS observed-frame spectrum of SD X-1.
Note the presence of a broad feature at $\sim 5600$ \r{A}, which we associate with the same feature identified as \ion{Mg}{II} in the spectrum of \protect\cite{Perlman98a}. Additionally, note the two narrower features, not found in the \protect\cite{Perlman98a}, at approximately the rest wavelengths of H$\alpha$ and H$\beta$. The rise in the continuum redward of 6500 \r{A} is instrumental, and the feature at $\sim 5600$ \r{A} is telluric.
}
\label{gmosspecx1}
\end{figure*}

The GMOS-S spectrum of this object, shown in Figure \ref{gmosspecx1}, is more complicated.
The doublet structure of the 5250 \r{A} feature is more clearly revealed. Additionally, two narrow absorption features centred at 6566 \r{A} and 4863 \r{A} also appear in this spectrum.
These lines have observed wavelengths consistent with H$\alpha$ and H$\beta$  at the systemic velocity of Sculptor Dwarf.
The line widths are marginally larger than the spectral resolution, though there is overlap with the measured width of the CuAr lamp spectra.
If these narrow absorption lines are associated with the same object as the BL Lac, then their presence is difficult to explain, as their rest-frame wavelengths are not associated with any known absorption features in AGN.
In order to understand the nature of this object, we have examined available catalogues with counterparts at other wavelengths.

\begin{figure*}
\centering
\includegraphics[width=0.8\textwidth]{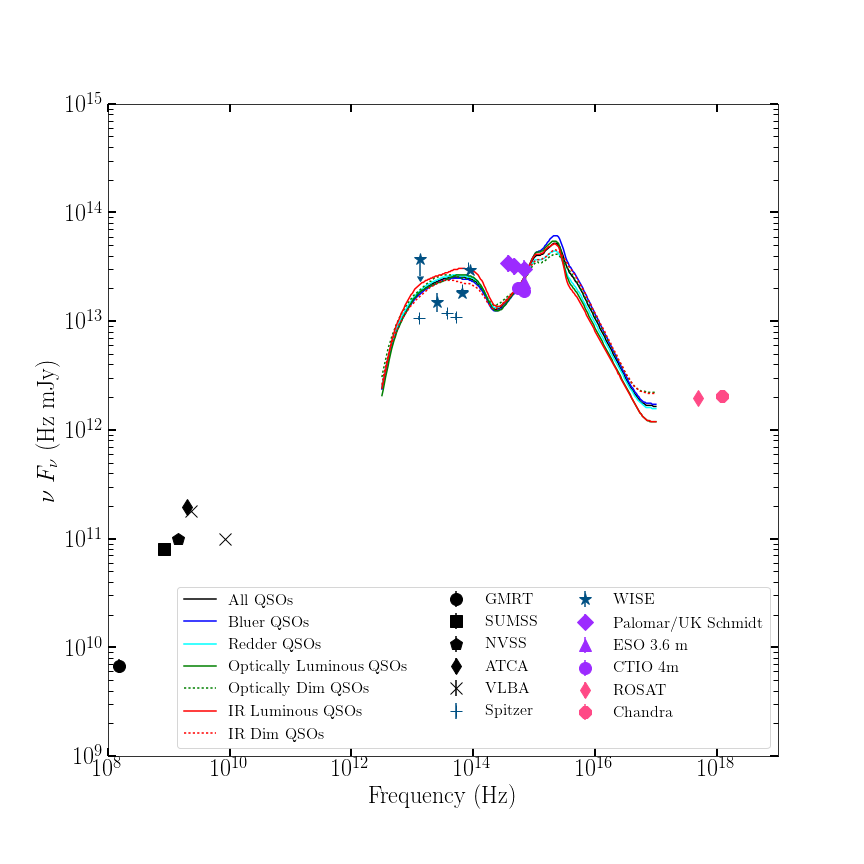}
\caption{%
Non-simultaneous spectral energy distribution (SED) for SD X-1 and all counterparts.
Error bars are included from catalogues where estimated, and upper limits are indicated when available.
Curves plotted over the SED are composite QSO spectra from \protect\cite{Richards06a}. Curves are normalized to have the same V-band flux density as the value measured by CTIO 4m V band photometry \protect\citep{Schweitzer95a}.
}
\label{SED}
\end{figure*}

\begin{figure*}
\includegraphics[width=0.45\textwidth]{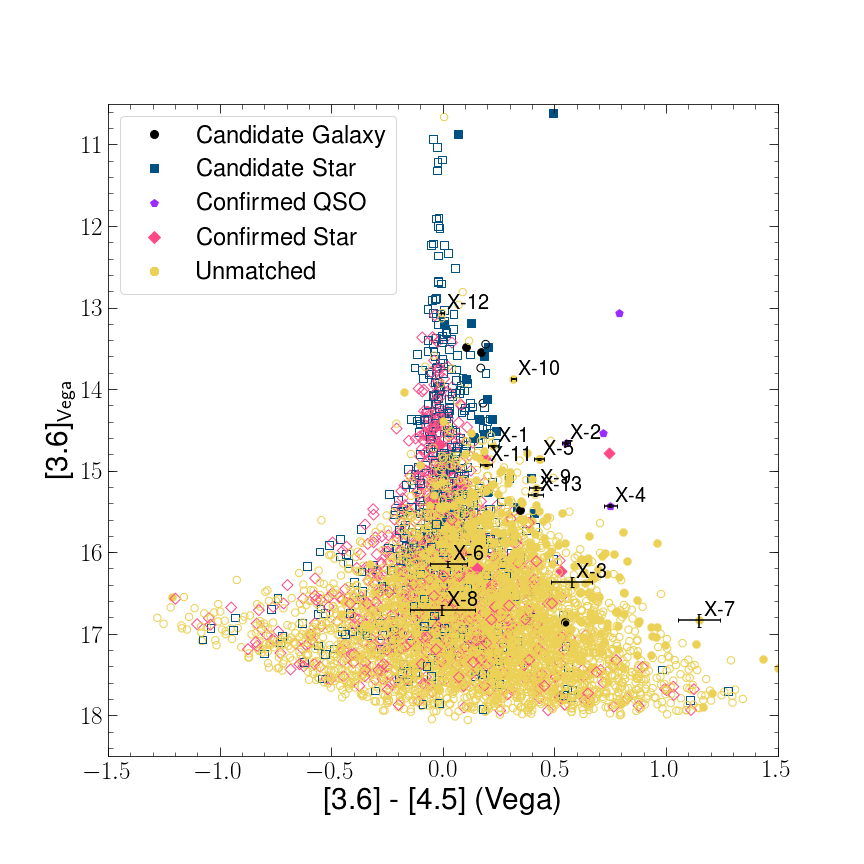}
\includegraphics[width=0.45\textwidth]{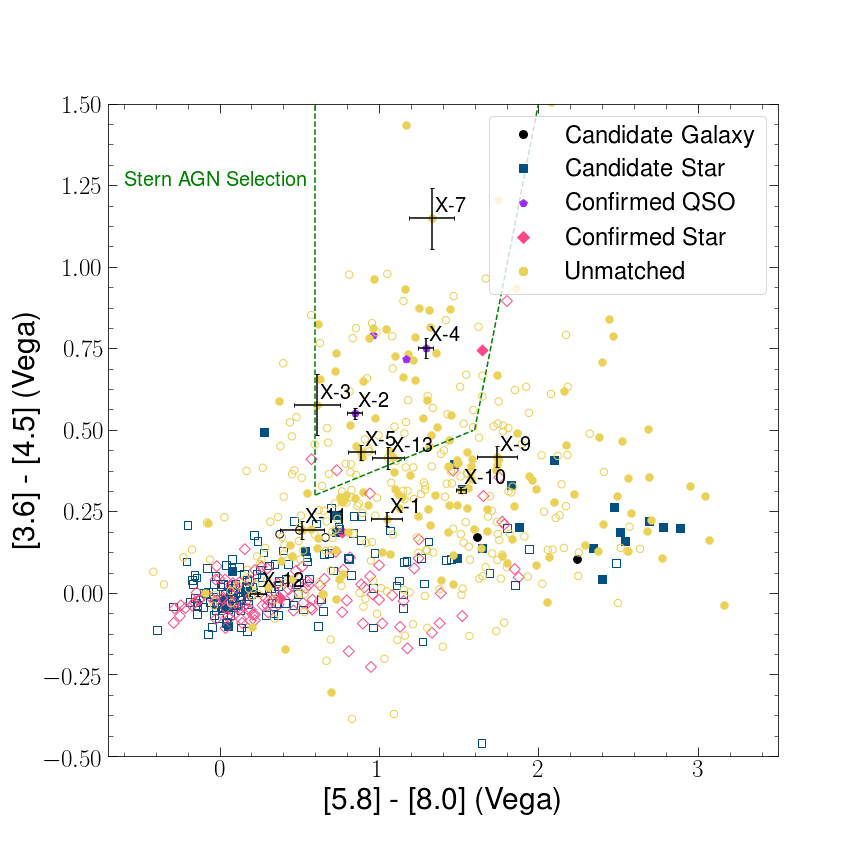}
%\hfill
\caption{%
Left: \Spitzer\ [3.6] $\mu$m vs [3.6] $\mu$m - [4.5] $\mu$m CMD for Sculptor.
Right: \Spitzer\ [3.6] $\mu$m - [4.5] $\mu$m vs [5.8] $\mu$m - [8.0] $\mu$m CCD for Sculptor.
We plot only sources with a [3.6] $\mu$m magnitude uncertainty  $< 0.2$. Sources with a MIPS 24.0 $\mu$m counterpart detected are filled, those without MIPS detections are unfilled.
Matches to X-ray sources are plotted with error bars.
We also plot the \protect\cite{Stern05a} AGN selection wedge.
Note that candidate and confirmed galaxies, stars and QSOs are based on matching to NED and SIMBAD.
}
\label{spitzercmd}
\end{figure*}

Aside from the initial radio and optical counterparts in QSO catalogues, there are substantial archival observations of the source, though none are simultaneous.
A spectral energy distribution (SED) of available archival data is plotted in Figure \ref{SED}.
Radio observations of the source made with the Australia Telescope Compact Array (ATCA, 2011 July/August) and the NRAO VLA Sky Survey (NVSS, 1996 Sept--Oct) seem to suggest time-variability in the radio flux measured at 1.4 and 2.0 GHz, respectively.
GALEX images do not show any counterpart at X-1's location.
Spitzer photometry reveals a very red object (see Figure \ref{spitzercmd}) that is very bright in redder IRAC channels (5.8 and 8.0 $\mu$m) relative other sources in the field, but imaging shows it is possibly contaminated by a nearby object at the longest wavelengths.
This source lies outside of the Stern AGN selection wedge, as shown in Figure \ref{spitzercmd}.
Ground based optical photometry gives the source $BVRI\sim 20$.

The most plausible explanation is that the observed source is a blend between the BL Lac object (contributing the broad optical emission lines, radio, IR, and X-ray emission) and a foreground object in Sculptor contributing the narrow absorption lines.
However, this explanation is clearly incomplete.
It does not provide a reason why the absorption lines were not observed in the previous ground-based spectrum, or their specific origin.
We also lack a plausible candidate for the source of the Balmer absorption lines if in Sculptor: the age of Sculptor's stellar populations would prohibit normal strong Balmer line sources, such as A stars.

One possibility for the origin of these lines is that they are absorption caused by a cloud of \ion{H}{i} gas in Sculptor along the line of sight to the AGN.
We can estimate the required column density using the curve of growth and assuming the cloud is optically thin. Using the measured equivalent widths of the lines, we then calculate the column density as:
\begin{equation}
N = 1.13\times 10^{20} \frac{EW}{\lambda^{2}f}\: \rm{cm^{-2}}
\end{equation}
where EW is the equivalent width in \r{A}, $\lambda$ is the rest wavelength of the line in \r{A}, and $f$ is the line oscillator strength \citep{Spitzer74,Frisch72}.
Using $f_{H\alpha} = 0.64$, $f_{H\beta} = 0.12$, and measuring $EW_{H\alpha} = 2.31 \pm 0.3$ \r{A} and $EW_{H\beta} = 2.02 \pm 0.2$ \r{A} from the lines, we derive column densities $N_{H\alpha} = 9.4 \times 10^{12}$~cm$^{-2}$ and $N_{H\beta} = 8.1 \times 10^{13}$ ~cm$^{-2}$.
These column densities represent the density of hydrogen atoms in the $n = 2$ state along the line of sight.
We can compare this to the neutral hydrogen density (presumed to be ground state hydrogen) observed by \cite{Bouchard03}.
The column density at the location of SD X-1 using Parkes to detect the 1.4~GHz spin-flip transition falls on a contour with $N_{n=1} = 1 \times 10^{18}$.
The ratio of the $N_{n=2}$ column density to the $N_{n=1}$ column density implies, assuming the Boltzmann distribution, temperatures of $\sim$20,000~K.
It is difficult to explain such a high ISM temperature inside Sculptor, especially given the dearth of potential ionizing sources in the galaxy that could heat the ISM.
However, since the ratio compares a spin-transition to an ordinary atomic electron transition, it is likely that the assumption of the Boltzmann distribution is inaccurate.
Ideally, the required column density of neutral hydrogen should be calculated using a measurement from the Lyman transitions.
As such, based on existing data, we cannot conclude whether these lines are due to intervening \ion{H}{i}  gas in Sculptor.

\begin{figure*}
\centering
\includegraphics[width=\textwidth]{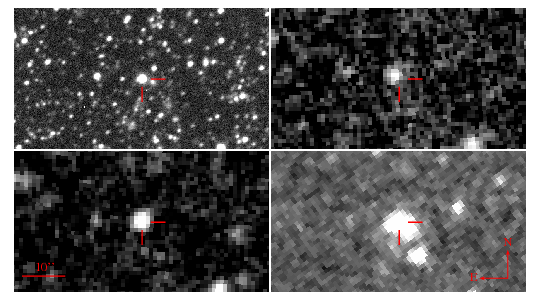}
\caption{%
Finding charts for SD X-1. Clockwise from left: Gemini GMOS-S r-band pre-imaging (2008), USNO AAOR image (1997), USNO SRCJ image (1982), and \Spitzer\ IRAC 3.6 $\mu$m image.
Up is North and left is East.
The X-ray position of SD X-1 is marked with a 1$\arcsec$ circle.
To the resolution of the GMOS pre-imaging, SD X-1's r-band counterpart appears to be a point-source.
Note the lack of any obvious high proper motion interlopers in the USNO imaging, while the \Spitzer\ image shows a nearby source which may be contaminating photometry at IR wavelengths.
}
\label{findingchartx1}
\end{figure*}

Another possibility for the origin of the Balmer lines is that they are caused by a star with high proper motion that lies in the slit of the GMOS spectrum but is outside of it in the \cite{Perlman98a} spectrum.
The GMOS pre-imaging of the source shows a single point source at the resolution of the observation (roughly 0.5\arcsec).
To look for a counterpart, we inspected archival USNO catalogues, shown with GMOS pre-imaging and \Spitzer\ imaging in Figure \ref{findingchartx1}.
The USNO images do not contain any plausible source that could be a high proper-motion object appearing in the slit of the GMOS spectrum.
We note, as a caveat, that these images are shallower and have poorer angular resolution than the GMOS imaging.
We also do not see any obvious interloper in the GMOS slit, as shown in Figure \ref{stamps1}.
Based on the GMOS spectrum, examination of the SED, and comparison with catalogues, we conclude that SD X-1 is most likely a background AGN with some foreground object in Sculptor causing foreground absorption, though the source of this absorption is unknown.

\subsection{SD X-2}
\begin{figure*}
\includegraphics[width=0.45\textwidth]{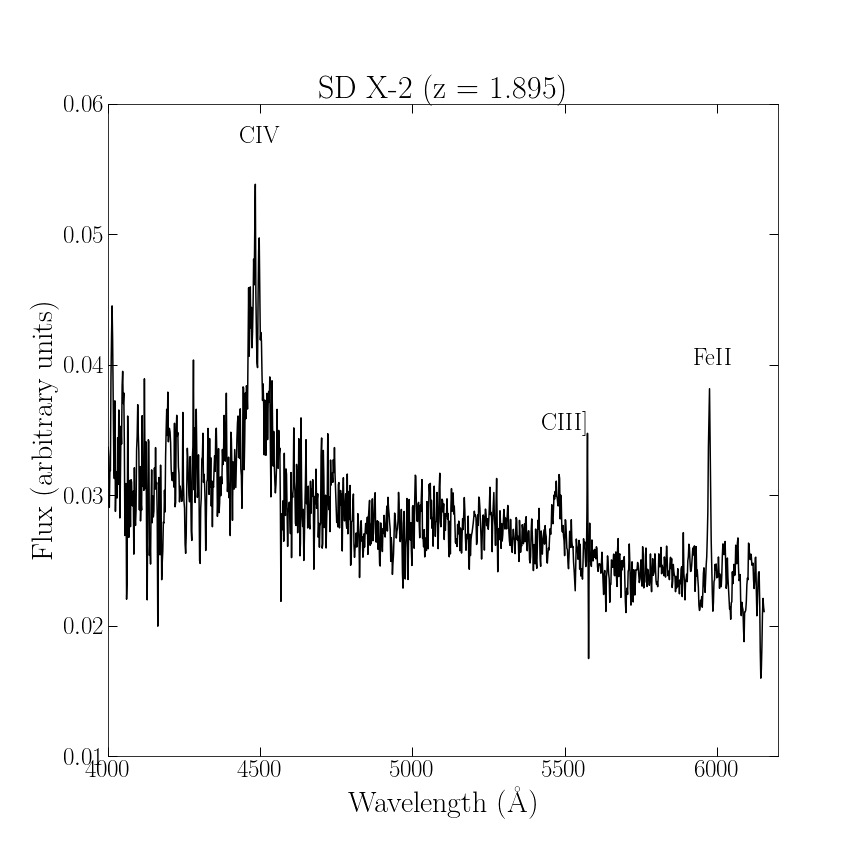}
\includegraphics[width=0.45\textwidth]{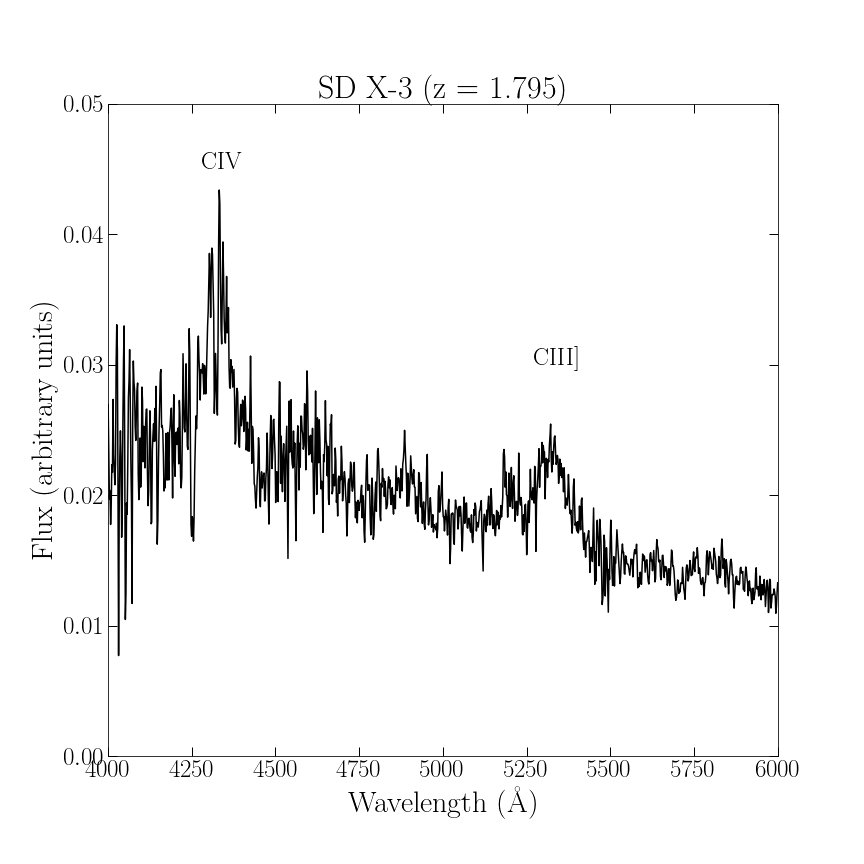}

\includegraphics[width=0.45\textwidth]{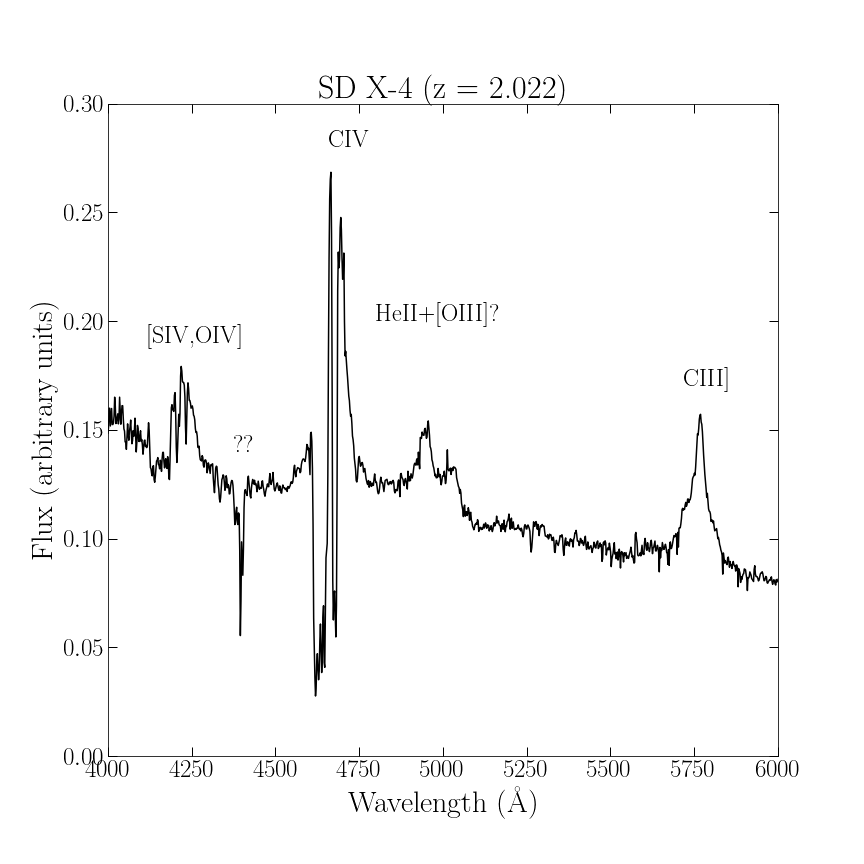}
%\hfill
\includegraphics[width=0.45\textwidth]{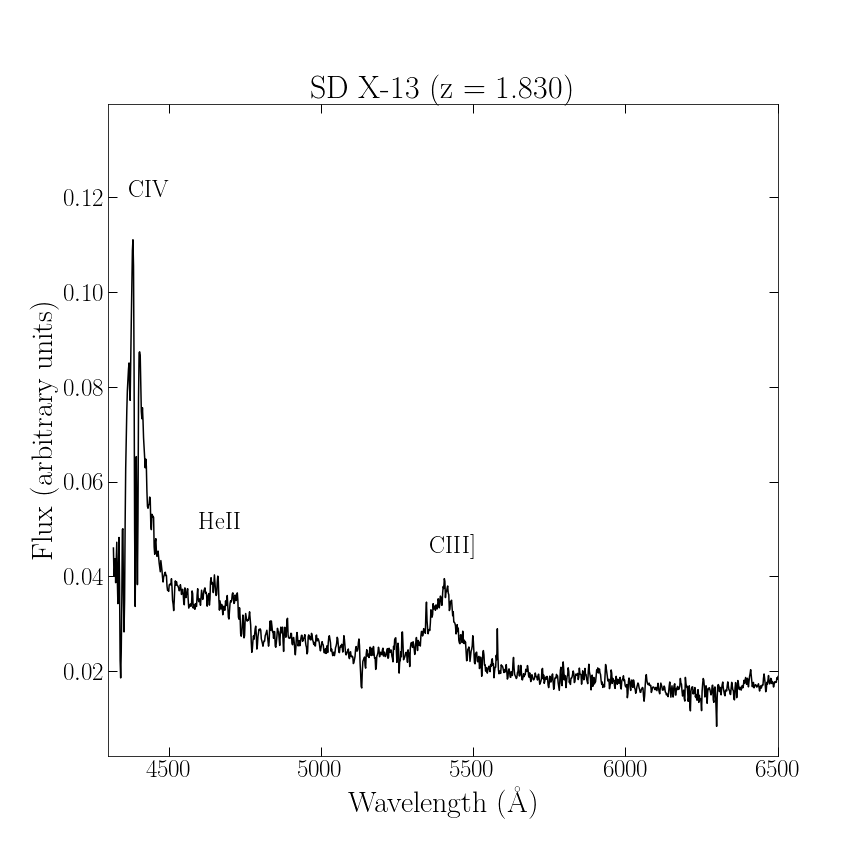}
\caption{%
GMOS-S MOS observed-frame spectra for SD X-2 (top left), SD X-3 (top right), SD X-4 (bottom left), and SD X-13 (bottom right).
The feature at $\sim$5600 \r{A} in the spectra of X-2 and X-13 is telluric.
}
\label{gmosspec}
\end{figure*}

This X-ray source also has an existing counterpart in the literature.
\cite{Tinney97a} identify the source as QJO100-3341 with an optical spectrum showing lines at $\sim$4500 \r{A}, $\sim$7900 \r{A}, and 8000 \r{A}.
They identify these lines as \ion{Mg}{ii}, H$\beta$ and [\ion{O}{iii}], with a calculated redshift of 0.602.
In our optical spectrum, shown in Figure \ref{gmosspec}, we detect an emission line at 4485 \r{A}, as well as new lines at 5476 and 5975 \r{A}.
We compare the observed-frame positions of these lines to strong lines in the SDSS template spectra. The positions of these new lines allow us to self-consistently identify the line at 4485 \r{A} as \ion{C}{iv}, making the other lines \ion{C}{iii}] and \ion{Fe}{ii}. From these identifications, we derive a new redshift of $z=1.895$.
Based on the broad lines in the optical spectrum, we classify this source as an AGN.

\subsection{SD X-3}
This X-ray source has no existing counterpart in the literature, aside from detections in photometric surveys.
We find that this source has an optical counterpart in our GMOS imaging with $r = 21.2$ and ${\rm H\alpha} = 21.1$.
Additionally, in our GMOS-MOS observation we have discovered a new spectroscopic counterpart.
The GMOS spectrum, shown in Figure \ref{gmosspec} contains two broad lines.
One line is at $\sim 4300$\r{A}, and the other is at at $\sim 5300$\r{A}.
Comparing with strong lines in SDSS template spectra, we can self-consistently assign these lines to be \ion{C}{iv} and \ion{C}{iii}]. With this tentative identification, X-3 has a derived redshift of $z=1.795$.
X-3's IR emission falls at the edge of the Stern AGN selection wedge.
The presence of broad lines in the optical spectrum leads us to classify this source as an AGN.

\subsection{SD X-4}
SD X-4 has an existing optical counterpart in the literature.
\cite{Tinney99a} obtained an optical spectrum showing a BAL QSO (ID: QJ0059-3344) with \ion{C}{iii}] and \ion{Mg}{ii} lines at $\sim$5900 and $\sim$8200 \r{A}, respectively, with a redshift of $z=2.022$.
An object at the same location was found by \cite{Jones09a}, but with a lower S/N spectrum and a number of weak lines identified with a redshift of $z= 0.07630$.
The GMOS spectrum of this object, shown in Figure \ref{gmosspec}, is of considerably higher quality than either existing spectrum in the literature in the region 4000--6000 \r{A}.
We identify the same \ion{C}{iii}] line and BAL QSO features as \cite{Tinney99a}, confirming the redshift of $z=2.022$.
In addition, we further identify \ion{C}{iv} as the absorbed feature at $\sim$4660 \r{A}, and find additional features at 4220 and 4800 \r{A}, which we identify as \ion{S}{iv},\ion{O}{iv}] and \ion{He}{ii}, respectively.
This object's IR counterpart falls deep in the AGN selection wedge, as shown in Figure \ref{spitzercmd}. We also identify a Gemini counterpart with $r = 19.4$ and ${\rm H}\alpha = 19.2$.
Based on the optical spectrum and its IR colours, we classify this object as an AGN.

\subsection{SD X-5}
The optical counterpart of SD X-5 was previously identified by \cite{Maccarone05a}, however it was not targeted for GMOS spectra as it has magnitude $V = 23.68$. With the Australia Telescope Compact Array (ATCA), \cite{Regis15a} identified a radio source $\sim2$\arcsec away that could potentially be associated.
\Spitzer\ matching identifies a counterpart which has colours in the Stern AGN region.
The $L_{X}$/$L_{\rm opt}$ ratio of this object is rather high compared to other AGNs in this field, and as such this object could potentially be a qLMXB or a dusty AGN.
Based on the IR colours, we tentatively classify this object as an AGN.

\subsection{SD X-6}
There is no existing literature identification for SD X-6.
This source is also not detected in the 2nd epoch of \Chandra\ observations.
It is detected in both Spitzer and Gemini GMOS imaging, though it lacks 5.8 $\mu$m and 8.0 $\mu$m detections.
Its position on the Spitzer CMD puts it closer to the main sequence than AGN or background galaxy counterparts.
In addition, its position on the r-H$\alpha$ CMD, shown in Figure \ref{gmoscmd}, suggests that it has less H$\alpha$ emission relative to ordinary sources in the galaxy.
It is difficult to make a definitive identification, given that this source has no followup X-ray detection, partial IRAC detection, and no spectrum.
However, given the H$\alpha$ dearth relative to the main sequence, it appears it is unlikely to be an XRB in Sculptor.
Possible classifications could be an AGN, an AGN blended with a foreground star in Sculptor, or a foreground MW star.

\subsection{SD X-7}
SD X-7 has a match in the quasar catalogue published by \cite{Flesch17a}. The catalogue entry lists $R$ (20.31) and $B$ (20.37) magnitudes.
However, the catalogue entry indicates that the B and R magnitudes were measured years apart.
Since QSOs can show variability, the colour is unlikely to be reliable.
The catalogue lists a probability of being a QSO $P_{QSO} = 0.98$, though it is unclear how the probability is derived from the optical photometry, and there is no spectroscopic classification of this object in the catalogue.
It was not targeted for a GMOS spectrum.
It lies outside of the field of view in the 2nd epoch of \Chandra\ observations, and was therefore not detected.
This source has a photometric counterpart detected by \Spitzer\ but is outside the field of view of our Gemini imaging.
SD X-7 has a counterpart in DES photometry with a separation of 0.075\arcsec, as shown in Figure \ref{DEScmd}.
In the bluer CMD, X-7 appears consistent with a horizontal branch star.
However, in the redder CMD, X-7 is slightly redder than the main sequence overall, suggestive of an AGN.
The IR counterpart is relatively faint and red, as can be seen in Figure \ref{spitzercmd}. Additionally, its position on the Spitzer colour-colour diagram puts it deep in the Stern AGN selection.
Based on its IR colours, we classify X-7 as a background AGN.

\subsection{SD X-8}
\label{sec:sdx8}
This source appears to be at a slightly different location from where it was reported by \cite{Maccarone05a}.
\texttt{wavdetect} finds that the coordinates of this source are $\sim2$\arcsec\ away from the first reported X-ray position.\footnote{%
This is not due to the combination of observations from multiple epochs: merging only the first epoch observations gives the same result. We believe there is a typographical error in the previous reported position.
}
We also note that this source is faint and it appears extended, possibly due to off-axis effects.
The 3$\sigma$ position found by \texttt{wavdetect} overlaps with a counterpart in \Spitzer\ imaging.
The position on the \Spitzer\ CMD suggests that it is an ordinary star, though it is not detected at 5.8 $\mu$m or 24.0 $\mu$m.
This mid-infrared counterpart corresponds with an optical counterpart located $\sim2$\arcsec\ away from the reported X-ray position, found by \cite{Kirby13a} and \cite{Kirby15a}.
We also find a Gemini GMOS counterpart to the \Spitzer\ object in $r$ and $H\alpha$, which has $r - {\rm H}\alpha$ colour consistent with the main sequence (no excess).
The \cite{Kirby13a} counterpart is a red giant with T$_{\rm eff} = 4904$~K and $\log(g) = 2.10$.
If this red giant were the Roche lobe-filling donor in an XRB with a 0.5 - 10 $M_{\odot}$ accretor, the system would have a period of $\sim$weeks.
This red giant was only observed once in the study of \cite{Kirby15a}, so we cannot look for radial velocity variation due to an unseen companion.
If the red giant is a companion in an XRB system, it is possible that the spectrum could show emission lines associated with XRBs, such as H$\alpha$.
However, the spectrum of this red giant from \cite{Kirby15a}, kindly provided to us by the authors, is not unusual.
In particular, it does not show H$\alpha$ emission, which might be expected of such a source (E. Kirby, private communication).
The lack of H$\alpha$ excess in either spectral or photometric measurements suggests that this red giant is unaffiliated with the X-ray source.
The optical spectrum of this source also shows no evidence of AGN emission, suggesting that it is not blended with an AGN.
The Dark Energy Survey catalogue identifies no counterpart except for a source consistent with location of the red giant, and no other counterpart is evident in the DES image in Figure \ref{x8stamp}.
Based on this, we conclude that the \Spitzer\, Gemini, and DES counterparts are the red giant from \cite{Kirby13a}, and are not associated with the X-ray source.
As such, we detect no optical/IR counterpart to X-8.
The Dark Energy Survey has a limiting magnitude of $\sim 24$ in $r$-band \citep{Abbott18a}, which should detect sources down to an absolute magnitude of $\sim 4.3$, assuming a distance modulus of 19.67 \citep{McConnachie12a}.
Therefore, based on the lack of an associated optical counterpart, we tentatively classify X-8 as a background AGN that is nearby on the sky to an unaffiliated red giant.

\subsection{SD X-9}
This source, previously classified as a background galaxy by \cite{Schweitzer95a}, lies out of the field of the 2nd epoch of \Chandra\ observations and was not detected.
It also lies out of the field for GMOS imaging.
It possesses an IR counterpart, with relatively red IR colours.
Based on prior identifications, we maintain the classification as a background galaxy.

\subsection{SD X-10}
SD X-10 is the brightest X--ray source in the second epoch of \Chandra\  observation.
This source lies outside of the field covered by the S3 chip in the first epoch of observations, however it is detected on other chips and is included in the X-ray source catalogue of \cite{Liu11a}.
It has no obvious literature counterparts, however its X-ray identification is complicated by \texttt{wavdetect} finding 2 sources in this location, spread out over a few arcseconds.
This could be a true extended object, an artifact of being off-axis, or multiple sources at the same location.
The source is not detected by Gemini, but has a \Spitzer\ counterpart 2.5$\arcsec$ away from the X-ray location.
This \Spitzer\ counterpart has a very red IR colour placing it outside the Stern AGN selection.
In the DSS and DES images of these coordinates, the latter of which is shown in Figure \ref{x10stamp}, we find a possibly extended source, spread over a few arcseconds, most likely a background galaxy.
Based on DSS, DES, and \Chandra\ both showing a bright extended object, we conclude that this source is a background galaxy.

\subsection{SD X-11}
SD X-11 has no known counterparts in the literature.
It has a relatively hard X-ray spectrum.
This source is also detected in the first epoch of \Chandra\ data, but it was not reported by \cite{Maccarone05a}, most likely due to a lack of an optical counterpart in the \cite{Schweitzer95a} catalog.
At Sculptor's distance, this object would have an X-ray luminosity of $\sim3.5 \times 10^{34}$~erg~s$^{-1}$, plausible for an LMXB.
It is detected in \Spitzer\ but not Gemini imaging.
As such, it was not targeted for GMOS spectroscopy.
Its \Spitzer\ colours lie outside the AGN selection wedge, however its colours are not consistent with those of an normal star.
Since the object is detected in X-ray and IR but has no optical counterpart, it could be a dusty AGN or a high redshift galaxy.
Based on the available photometry and X-ray properties, we identify this source as a candidate AGN/background galaxy.

\subsection{SD X-12}
\begin{figure*}
\includegraphics[width=0.9\paperwidth]{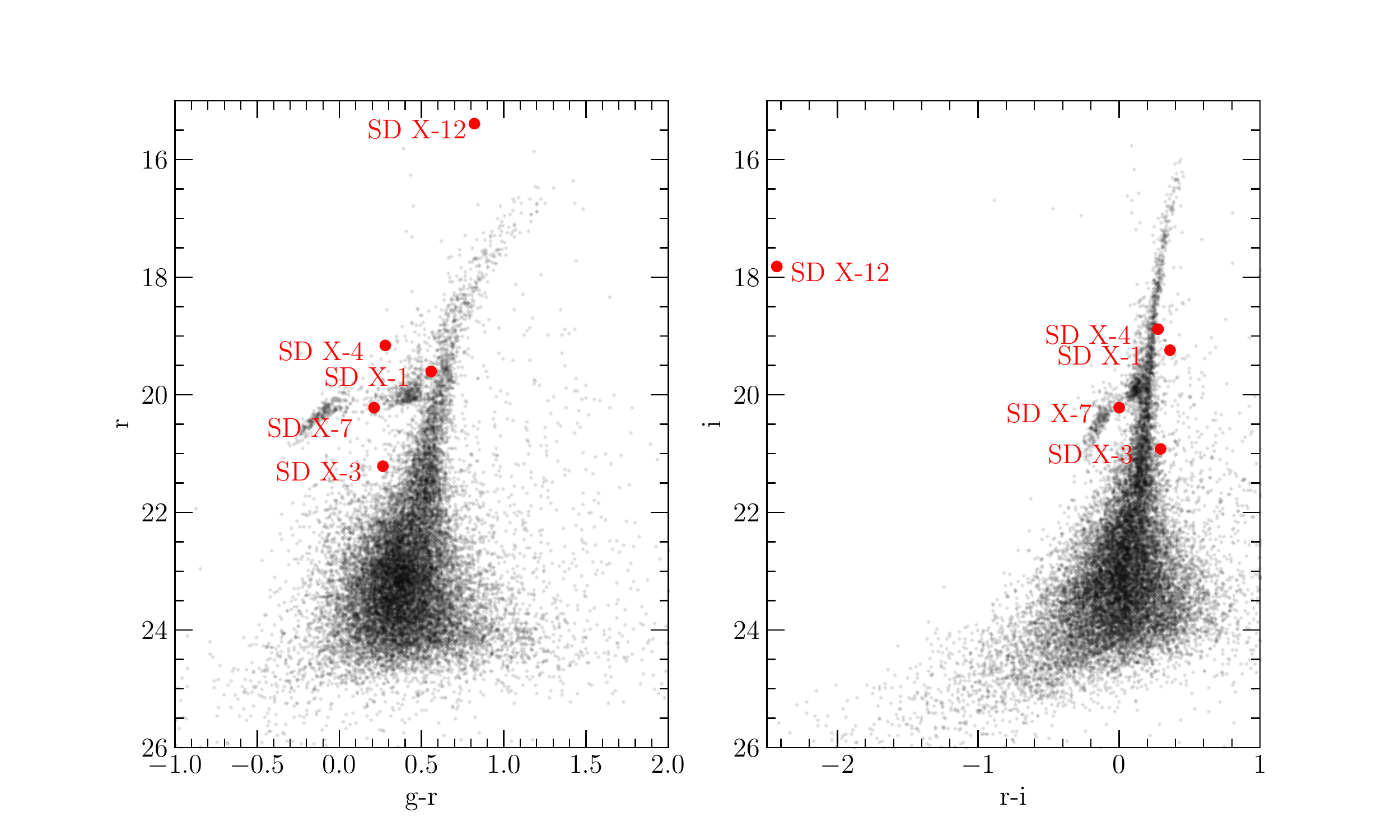}
\caption{Dark Energy Survey (DES) colour-magnitude diagrams (CMDs) for the Sculptor Dwarf Galaxy.
Note that all counterparts to X-ray sources except X-7 do not appear to have a consistent position relative to Sculptor's stellar population, and X-12 does not look like a Sculptor member at all.}
\label{DEScmd}
\end{figure*}

SD X-12 has no known counterparts in the literature.
It is detected in the \Spitzer\ and Gemini imaging, however it is saturated in Gemini.
We also identify a very bright counterpart in DES with a separation of 0.0075\arcsec, shown in Figure \ref{x12stamp}.
The IR counterpart has $[3.6]_{\rm Vega}$ = 13.07, making it one of the brightest objects in the population.
This object was detected in the first epoch of \Chandra\ data, but was not identified by \cite{Maccarone05a}, most likely because it was not in the \cite{Schweitzer95a} catalogue due to saturation.
Its position in the colour-colour diagram (see Figure \ref{spitzercmd}) is in a portion of the CCD occupied primarily by confirmed or candidate stars in Sculptor.
However, the photometry of this counterpart revealed by DES is clearly distinct from the ordinary stellar population of Sculptor.
The object is several magnitudes brighter than giant branch stars in Sculptor, as seen in the bluer CMD (see Figure \ref{DEScmd} left panel).
However, in redder filters, X-12's counterpart appears significantly redder than the ordinary population (see Figure \ref{DEScmd} right panel).
The overall brightness of this object and its large colour disparity from Sculptor's population suggests that it is likely to be a foreground star.

This object is unlikely to be an XRB or a CV, as the required optical contribution of the accretion disk creates an implausible optical/X-ray flux ratio given an implied X-ray luminosity of $\sim3.5 \times 10^{34}$~erg~s$^{-1}$ if it is in Sculptor. Additionally, this object is observed to show significant variability - the \textit{i} magnitude in the DES survey is $\sim2$ magnitudes fainter than the I magnitude measured in the USNO-B1 survey.
This object also has observed flaring in the ASASSN survey \citep{Shappee14a,Kochanek17a}.
The bright optical counterpart is also catalogued in Gaia DR 2 (source id: 5027218233097636736, \citealt{GaiaDR2}), with a confident parallax estimate of $1.45\pm0.04$~ mas clearly showing this source to be a foreground star in the solar neighborhood.
Based on the DES magnitudes, the presence of variability, and a significant Gaia parallax, we conclude that this source is a foreground active binary (AB) or flaring star.

\subsection{SD X-13}
\begin{figure}
\includegraphics[width=0.45\textwidth]{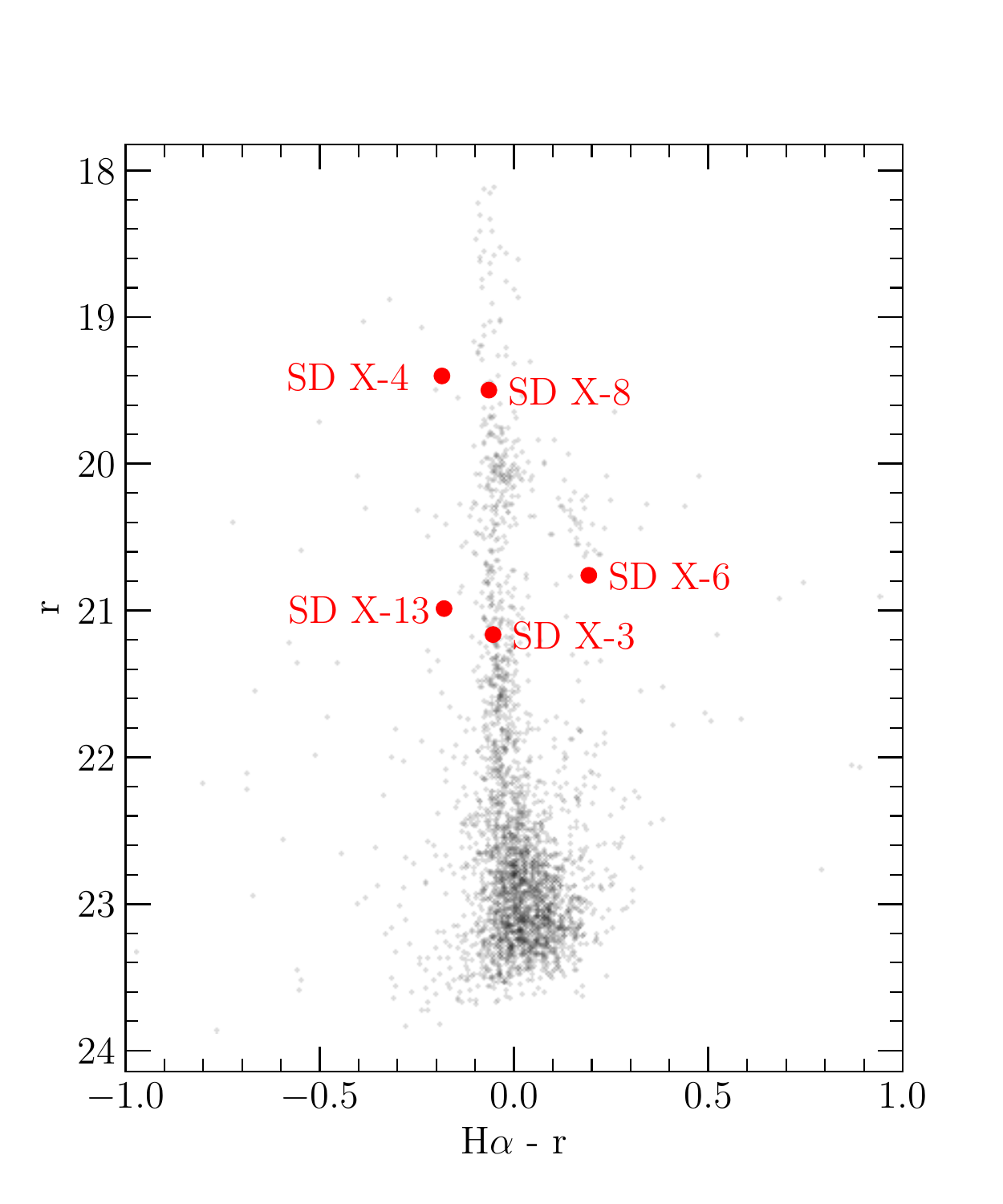}
\caption{%
Gemini GMOS r-H$\alpha$ CMD for the core ($5.5$ arcmin$^{2}$) of Sculptor.
X-ray sources with Gemini imaging counterparts are labelled.
Accreting sources are typically expected to show H$\alpha$ excess relative to the main sequence, which would be on the left side of the CMD.}
\label{gmoscmd}
\end{figure}

SD X-13 was found by WAVDETECT in the second epoch of \Chandra\ observations.
This object was located well off-axis on the S3 chip in the first epoch of observations, and so was not reported by \cite{Maccarone05a}, as its position could not be easily determined.
X-13 has IR colours within the Stern AGN wedge.
This object is also detected in our Gemini imaging, and is shown in Figure \ref{gmoscmd} to have an H$\alpha$ excess relative to the main sequence.
Additionally, the GMOS spectrum, shown in Figure \ref{gmosspec}, shows three broad emission features:
a strong, broad line that overlaps with the instrumental cutoff at $\sim$4400 \r{A}, a broad, faint double centered around 4600 \r{A}, and a strong, broad line at 5407 \rm{A}.
Through comparison with template spectra and typical strong lines, we conclude that the features at 4400, 4600, and 5407 \r{A} are \ion{C}{iv}, \ion{He}{ii}, and \ion{C}{iii}, respectively.
From this, we derive a redshift of $z=1.830$ and conclude that this source is a background AGN.

\begin{table*}

\caption{%
Summary of X-ray positions and literature identifications for the X-ray sources in the direction of Sculptor.
}
\label{sourcetablepos}
%\tabletypesize{\scriptsize}
\begin{adjustwidth}{-.5in}{-.5in}
\begin{tabular}{lccccc}
\hline \hline
Object & RA$_X$ & DEC$_X$ &
Other ID & Lit. Classification & Classification (TW) \\
 & & & & & \\
\hline
SD X-1 & 01:00:09.39 & -33:37:31.900 & PKS 0057-338\textsuperscript{[1]};WGAJ0100.1-3337\textsuperscript{[2]} & BL Lac & BL Lac + foreground \\
SD X-2 & 01:00:26.19 & -33:41:07.500 & QJO100-3341\textsuperscript{[1]}\textsuperscript{[3]} & AGN & AGN \\
SD X-3 & 00:59:58.68 & -33:43:37.100 & \nodata & \nodata & QSO \\
SD X-4 & 00:59:52.75 & -33:44:26.100 & QJ0059-3344\textsuperscript{[4]} & QSO & BAL QSO \\
SD X-5 & 01:00:13.84 & -33:44:43.300 & J010014.0-334442.7\textsuperscript{[5]} & \nodata & AGN Candidate \\
SD X-6 & 01:00:03.03 & -33:44:26.900 & \nodata & \nodata & AGN/AGN+FG/FG(MW) \\
SD X-7 & 00:59:59.89 & -33:38:12.200 & J005959.90-333811.4\textsuperscript{[6]} & QSO & AGN \\
SD X-8 & 00:59:54.38 & -33:44:30.460 & Scl 1008920\textsuperscript{[7][8][9]};J005954.2-334429\textsuperscript{[10]} & Red Giant & Unassociated Red Giant + AGN \\
SD X-9 & 01:00:29.22 & -33:47:37.100 & \nodata & Background Galaxy\textsuperscript{[11]} & Background Galaxy \\
SD X-10 & 01:00:27.33 & -33:51:10.345 & \nodata & \nodata & Background Galaxy \\
SD X-11 & 00:59:57.18 & -33:44:19.093 & \nodata & \nodata & AGN/Background Galaxy Candidate \\
SD X-12 & 01:00:02.85 & -33:43:04.166 & \nodata & \nodata & Flaring foreground star/AB \\
SD X-13 & 00:59:47.26 & -33:43:07.106 & \nodata & \nodata & AGN \\
\end{tabular}
\end{adjustwidth}
\bigskip
For sources re-detected or newly detected in the 2nd \Chandra\ epoch, we use the location determined from WAVDETECT in that observation.
Otherwise, we use the position reported in \protect\cite{Maccarone05a}.
[1]: \protect\cite{Tinney97a},[2]: \protect\cite{Perlman98a},[3]: \protect\cite{Veron10a},[4]: \protect\cite{Tinney99a},[5]: \protect\cite{Regis15a},[6]: \protect\cite{Flesch17a},[7]: \protect\cite{Kirby15a},[8]: \protect\cite{SIMBAD1a},[9]: \protect\cite{Walker09a},[10]: \protect\cite{Flesch16a},[11]:
\protect\cite{Schweitzer95a}

\end{table*}

\begin{table*}
\caption{Summary of X-ray properties for sources in the Sculptor Dwarf field.
\label{sourcetablexray}
}
\begin{tabular}{lcrrr}
\hline \hline
Object &CSC ID & Flux & HR1 & HR2 \\
 & & $10^{-14}$ erg s$^{-1}$ & & \\
\hline
SD X-1 & J010009.3-333732 & $6^{+1}_{-1}$ & $-0.3^{+0.1}_{-0.1}$ & $0.0^{+0.1}_{-0.1}$ \\
SD X-2 & J010026.1-334107 & $2.4^{+0.1}_{-0.1}$ & $-0.15^{+0.08}_{-0.08}$ & $-0.05^{+0.07}_{-0.07}$ \\
SD X-3 & J005958.6-334337 & $1.4^{+0.1}_{-0.1}$ & $-0.4^{+0.08}_{-0.08}$ & $-0.02^{+0.07}_{-0.06}$ \\
SD X-4 & J005952.7-334425 & $1.4^{+0.2}_{-0.2}$ & $0.08^{+0.08}_{-0.08}$ & $0.1^{+0.1}_{-0.1}$ \\
SD X-5 & J010013.9-334442 & $5^{+1}_{-1}$ & $-0.1^{+0.2}_{-0.2}$ & $-0.1^{+0.2}_{-0.2}$ \\
SD X-6 & \nodata & \nodata & \nodata & \nodata \\
SD X-7 & J005959.8-333812 & $1.8^{+0.2}_{-0.2}$ & $-0.4^{+0.1}_{-0.1}$ & $0.07^{+0.07}_{-0.07}$ \\
SD X-8 & J005954.3-334429 & $0.4^{+0.1}_{-0.1}$ & $0.2^{+0.2}_{-0.2}$ & $-0.0^{+0.2}_{-0.2}$ \\
SD X-9 & J010029.1-334735 & $5.3^{+0.3}_{-0.3}$ & $-0.2^{+0.1}_{-0.1}$ & $-0.14^{+0.06}_{-0.06}$ \\
SD X-10 & J010027.4-335111 & $36.8^{+0.6}_{-0.6}$ & $-0.22^{+0.03}_{-0.03}$ & $0.01^{+0.03}_{-0.02}$ \\
SD X-11 & J005957.1-334419 & $3.1^{+0.1}_{-0.1}$ & $0.07^{+0.06}_{-0.06}$ & $0.29^{+0.07}_{-0.08}$ \\
SD X-12 & J010002.8-334304 & $2.47^{+0.09}_{-0.09}$ & $-0.46^{+0.04}_{-0.04}$ & $-0.06^{+0.03}_{-0.03}$ \\
SD X-13 & J005947.2-334307 & $2.7^{+0.3}_{-0.3}$ & $-0.1^{+0.1}_{-0.1}$ & $-0.0^{+0.1}_{-0.1}$ \\
\end{tabular}

\bigskip

Fluxes and hardness ratios are taken from the corresponding source in the \textit{Chandra Source Catalog} (CSC). Flux is the 0.5--7.0 keV flux of the CSC source.
HR1 is the ``soft'' CSC flux hardness ratio defined as $(F_{2.0 - 7.0 \rm \ keV} - F_{1.2 - 2.0 \rm \ keV})/F_{0.5 - 7.0 \rm \ keV}$.
HR2 is the ``hard'' CSC flux hardness ratio defined as $(F_{1.2 - 2.0 \rm \ keV} - F_{0.5 - 1.2 \rm \ keV})/F_{0.5 - 7.0 \rm \ keV}$.

\end{table*}

\renewcommand{\arraystretch}{1}

\begin{table*}
\caption{IR/visible properties of SD X-ray sources.\label{sourcetableiropt} }
\begin{tabular}{lccccccccc}
\hline \hline
Object &
$[3.6]_{\rm Vega}$ & $[4.5]_{\rm Vega}$ & $[5.8]_{\rm Vega}$ & $[8.0]_{\rm Vega}$ &
$[24.0]_{\rm Vega}$ & r$_{\rm Vega}$ & H$\alpha_{\rm Vega}$ & X-ray/IR Offset \\
 & & & & & & & & \arcsec \\
\hline
SD X-1 & $14.70 \pm 0.01$ & $14.47 \pm 0.02$ & $14.33 \pm 0.08$ & $13.29 \pm 0.06$ & $9.82 \pm 0.05$ & \nodata & \nodata & 1.04 \\
SD X-2 & $14.67 \pm 0.01$ & $14.12 \pm 0.01$ & $13.34 \pm 0.04$ & $12.50 \pm 0.03$ & $8.63 \pm 0.02$ & \nodata & \nodata & 0.45 \\
SD X-3 & $16.37 \pm 0.06$ & $15.79 \pm 0.07$ & $14.5 \pm 0.1$ & $13.9 \pm 0.1$ & $9.84 \pm 0.05$ & $21.164 \pm 0.006$ & $21.11 \pm 0.02$ & 0.16 \\
SD X-4 & $15.43 \pm 0.02$ & $14.68 \pm 0.02$ & $13.66 \pm 0.04$ & $12.37 \pm 0.03$ & $8.82 \pm 0.02$ & $19.401 \pm 0.001$ & $19.216 \pm 0.004$ & 0.16 \\
SD X-5 & $14.85 \pm 0.01$ & $14.42 \pm 0.02$ & $14.07 \pm 0.06$ & $13.18 \pm 0.05$ & $10.12 \pm 0.06$ & \nodata & \nodata & 1.54 \\
SD X-6 & $16.14 \pm 0.04$ & $16.12 \pm 0.07$ & \nodata & \nodata & $10.38 \pm 0.07$ & $20.760 \pm 0.004$ & $20.95 \pm 0.02$ & 1.7 \\
SD X-7 & $16.84 \pm 0.08$ & $15.69 \pm 0.05$ & $14.7 \pm 0.1$ & $13.42 \pm 0.07$ & $10.51 \pm 0.07$ & \nodata & \nodata & 0.12 \\
SD X-8 & $16.71 \pm 0.06$ & $16.7 \pm 0.1$ & \nodata & $15.4 \pm 0.4$ & \nodata & $19.498 \pm 0.002$ & $19.433 \pm 0.005$ & 2.14 \\
SD X-9 & $15.21 \pm 0.02$ & $14.80 \pm 0.03$ & $14.7 \pm 0.1$ & $12.95 \pm 0.04$ & $9.52 \pm 0.04$ & \nodata & \nodata & 1.05 \\
SD X-10 & $13.874 \pm 0.006$ & $13.558 \pm 0.008$ & $13.20 \pm 0.03$ & $11.69 \pm 0.01$ & $7.79 \pm 0.02$ & \nodata & \nodata & 2.52 \\
SD X-11 & $14.93 \pm 0.01$ & $14.74 \pm 0.02$ & $14.39 \pm 0.09$ & $13.9 \pm 0.1$ & $10.42 \pm 0.08$ & \nodata & \nodata & 0.05 \\
SD X-12 & $13.068 \pm 0.004$ & $13.071 \pm 0.006$ & $13.18 \pm 0.03$ & $12.94 \pm 0.04$ & \nodata & \nodata & \nodata & 0.21 \\
SD X-13 & $15.29 \pm 0.02$ & $14.88 \pm 0.03$ & $14.35 \pm 0.08$ & $13.29 \pm 0.06$ & \nodata & $20.987 \pm 0.005$ & $20.81 \pm 0.01$ & 0.84 \\
\end{tabular}

\bigskip

X-ray/IR Offset was determined by matching the \Chandra\ coordinates to the \Spitzer\ catalogue using TOPCAT.
Note that $H\alpha$ is calibrated relative such that the main sequence has ${\rm H}\alpha - r = 0$, rather than having an absolute calibration.

\end{table*}

\section{Discussion}
In this study, we have completed the deepest, most thorough survey of an isolated, old stellar population with a low stellar encounter rate, and we have found that amongst the bright X-ray sources in Sculptor's direction, 8 are AGNs or AGN candidates, 2 are background galaxies, 1 is a foreground star, and 2 have uncertain identifications.
In the conservative case, Sculptor appears to lack any bright XRBs in the present day.
Our examination of the bright X-ray source population in Sculptor shows no strong XRB candidates.
In addition, the X-ray luminosity function for this galaxy, shown in Figure~\ref{lognlogs}, shows no excess above what is expected due to background sources from deep field studies, suggesting an absence of faint XRB candidates.
\cite{Dehnen06a} used the previous result of \cite{Maccarone05a} to argue that Sculptor may need a dark matter halo of $10^{9}$ M$_{\sun}$ in order to retain LMXBs, unless there exists a class of LMXBs with preferentially lower natal kicks.
Although low natal kick sources have been observed, these do not represent a large fraction of the LMXB population \citep{Podsiadlowski05a}.
However, the absence of bright LMXBs in Sculptor would tend to imply that such large dark matter halos are unnecessary.
If Sculptor is representative of dwarf galaxies in terms of present day XRB population, then there are also possible implications for interactions between a dwarf galaxy and its host or a dwarf galaxy and its globular cluster system.
First, one immediate implication is that dwarf galaxies do not contaminate their host galaxies with significant quantities of XRBs when they interact with them.
Additionally, if globular clusters can contaminate their host galaxy with XRBs, then Sculptor's lack of present day XRBs may suggest that it has also lacked globular clusters in the past.

\subsection{Primordial binary contributions to observed populations}
Dwarf galaxies like Sculptor can be shown to have very low stellar encounter rates.
Sculptor's radial brightness profile can be reasonably described using a King model \citep{Irwin95a}.
As such, we estimate the stellar encounter rate $\Gamma$ using the following relation \citep{Verbunt87a}:
\begin{equation}
\Gamma = \frac{\rho_{c}^{2} r_{c}^3}{\sigma}
\end{equation}
where $\rho_{c}$ is the central luminosity density, $r_{c}$ is the core radius, and $\sigma$ is the central velocity dispersion.
We use $r_{c} = 145$~pc, based on a distance of 86~kpc and an apparent core radius size of 5.8\arcmin \citep{Irwin95a,McConnachie12a}.
Additionally, we use $\sigma = 9.2$~km~s$^{-1}$ \citep{Burkert15a}.
We use central surface brightness $\mu_{c} = 10.6 L_{\sun}/\rm pc^{2}$, which corresponds to $\rho_{c} = 0.041 L_{\sun}/\rm pc^{3}$ using the relationbetween projected and volume luminosity densities given by \cite{Djorgovski93a}.
Under these assumptions, and using the normalization that 47 Tuc has $\Gamma$ = 1000, Sculptor dSph has $\Gamma = 0.009$.
Based on this analysis, we expect that any XRBs inside of Sculptor should be formed primordially rather than through stellar encounters.
Additionally, Sculptor's relative isolation and lack of globular clusters suggests that it is unlikely to have captured XRBs through galaxy-scale interactions with the MW or GCs.
Comparing with the calculations of $\Gamma$ to those of \cite{Bahramian13a} (Table 4), we find that Sculptor dSph has a very small stellar encounter rate compared to Galactic GCs, most similar to that of the low-density clusters Arp 2 and Palomar 4.
In general, these clusters tend to be distant, sparse, and low-mass compared to the overall population of galactic GCs, making them difficult to observe.
With a stellar mass of $2.3 \times 10^{6} M_{\sun}$, Sculptor outweighs many globular clusters by $\sim$ 1-2 orders of magnitude, though its total mass including dark matter is closer to $3 \times 10^{7} M_{\sun}$  \citep{McConnachie12a,Battaglia08a,Kimmig15a}.

From our study, we see that Sculptor is essentially devoid of XRBs in the present day, which implies that it has no bright primordial binaries that have survived to the present epoch.
There are a few explanations for why this may be the case.
The first is that, despite being embedded in large DM haloes, natal kicks to XRB systems may be ejecting them from dwarf galaxies.
\cite{Dehnen06a} investigated the size of dark matter halo needed to retain XRBs based on the initial reported discovery of LMXBs inside Sculptor, with the assumption that the dark matter within the visible galaxy is $5 \times 10^{7}$ M$_{\sun}$.
We can estimate the central escape velocity from Sculptor using this assumption and the following relation \citep{Dehnen06a}:
\begin{equation}
v_{esc}^2(r) = v_{0}^2 ln \frac{r_t^2 + r_c^2}{r^2 + r_c^2}
\end{equation}
where $r_c$ is the core radius, $r_t$ is the tidal radius, and $v_0$ is the asymptotic circular speed defined by:
\begin{equation}
v_0 = 12 \rm km s^{-1} \sqrt{ 1 + (\frac{r_c}{1.5 \rm kpc}) }
\end{equation}
We use $r_t = 15$~kpc and $r_c = 101$~pc to arrive at $v_{esc,0} \sim 38$~km~s$^{-1}$ \citep{Dehnen06a,Irwin95a}.
Although this value is relatively low, it is comparable to the central escape velocities of a number of Galactic globular clusters (see, for example, \cite{McLaughlin05a}.

This order of magnitude estimate suggests that Sculptor could retain primordial binaries in its core in the same manner as a globular cluster.
Therefore, the absence of bright XRBs in Sculptor would imply that the bright XRBs observed in GCs are all dynamically formed closer to the present epoch rather than primordial.
The second explanation for the lack of primordial bright binaries in Sculptor is that they were never created in the first place: Sculptor's low stellar mass means that not enough primordial XRBs were created to result in some surviving to the present day.
This in turn would imply that the minimum required mass for an old, isolated population to still have bright XRBs in the present day should be larger than a few $10^{6} M_{\sun}$.
As such, we can expect that Galactic globular clusters with an encounter rate similar to Sculptor should also be devoid of binaries, as they have much lower mass and a negligible stellar encounter rate.

\subsection{Implications for Dwarf Galaxies}
If Sculptor is representative of other dwarf galaxies in the Local Group this would imply that a large fraction of the X-ray sources in the field of nearby dwarfs are in fact unrelated to the galaxies themselves.
Many of these sources are likely to be either background AGN or foreground active/flaring stars.
Our search for X-ray sources suitable for spectroscopy places constraints on the X-ray luminosity function above a few $10^{34}$ erg~s$^{-1}$, and in the conservative case Sculptor appears to lack any sources brighter than this limit.
Two recent surveys of Draco dSph with \textit{XMM-Newton} each identified a handful of XRB candidates.
\cite{Sonbas16a} and \cite{Saeedi16a} each identify four candidate XRBs.
Three of these appear to be faint sources, possibly CVs, symbiotic stars, or qLMXBs. One source has a reported $L_{X} = 8\times10^{34}$~erg~s$^{-1}$, brighter than the limit we have investigated for Sculptor.
\cite{Nucita13a} investigated Fornax dSph with \textit{XMM-Newton}, finding in general that the number of X-ray sources was consistent with the predicted number of background sources for the area surveyed.
However, they also identified two sources potentially associated with globular clusters bound to Fornax.
This result appears to be consistent with Sculptor, which lacks globular clusters of its own.
\cite{Manni15a} investigated four dSphs (Draco, Ursa Major II, Ursa Minor, and Leo I), also finding that the number of sources detected in the direction of each galaxy was consistent with background predictions, but noting that they could not rule out the possibility of a limited number of these sources being associated with the galaxy itself.
It is interesting to note that a number of compact object-related phenomena are found preferentially in dwarf galaxies.
The only repeating fast radio burst, FRB 121102, has been localized to a dwarf galaxy \citep{Tendulkar17} and
both superluminous supernovae and long-duration gamma bursts seem to be preferentially located in dwarf galaxies \citep{Perley16a,Fruchter06a}.
It is therefore curious that dwarf galaxies in the local universe, like Sculptor, do not appear to have many compact objects.
FRB host galaxies do share a number of differences from DGs like Sculptor, the most prominent being that the host galaxy of FRB 121102 has a relatively high star formation rate of 0.4~$M_{\sun}$~yr$^{-1}$.
This suggests that properties like active star formation may be crucial for hosting exotic compact object phenomena or, more generally, unknown phenomena with proposed compact object origins.

\subsection{Future Studies}
Dwarf galaxies like Sculptor present a unique challenge for XRB searches because of their large angular size on the sky, which means that the expected number of AGNs contaminating any dwarf galaxy is expected to be large.
As such, the X-ray source population of dwarf galaxies can only be accurately characterized using deep, high-resolution multiwavelength observations, like the \Chandra, \Spitzer, and Gemini observations used in this study.

Although we have characterized the bright X-ray sources in this galaxy, limits on the optical depth of the study have prevented us from accurately characterizing the CV population.
A thorough study of Sculptor would require depths similar to those used in globular cluster surveys, so that the population of CVs can be identified and studied in detail.
For example, using the CV populations of M80 and NGC 6397 \citep{Pietrukowicz09a,Cohn10a} as analogues suggest that observations would require a depth of $R \sim 24.5$ to capture 50\% of the CV population in Sculptor.
In particular, CVs are often separable from the ordinary stellar population using UV or deep H$\alpha$ photometry.
Our H$\alpha$ limits are not deep enough to capture the typical CV population, and Sculptor lacks UV observations with \textit{HST}.
A future study of Sculptor's X-ray population would seek to characterize not only the bright population of X-ray sources, but also the fainter X-ray sources contained within Sculptor's population.

\section*{Acknowledgements}

R. M. A. acknowledges support from an NSERC CGS-D scholarship.
P. B. acknowledges support from an NSERC Discovery Grant.
S. Z. acknowledges funding from \Chandra\ X-ray Center grant GO8-9090X.
We thank S. Gallagher, J. Cami, M. Gorski, C. Heinke, S. Koposov, H. Cohn, and P. Lugger for helpful discussions.
We thank the anonymous referee for providing helpful suggestions for improving the manuscript.
We also thank E. Kirby for kindly providing the spectrum of Scl 1008920 and P. Stetson and C. Gallart for kindly providing reference photometry of Sculptor.
This work is based in part on observations made with the Spitzer Space Telescope, which is operated by the Jet Propulsion Laboratory, California Institute of Technology under a contract with NASA.
Based on observations obtained at the Gemini Observatory, acquired through the Gemini Observatory Archive and processed using the Gemini python IRAF package, which is operated by the Association of Universities for Research in Astronomy, Inc., under a cooperative agreement with the NSF on behalf of the Gemini partnership: the National Science Foundation (United States), the National Research Council (Canada), CONICYT (Chile), Ministerio de Ciencia, Tecnolog\'{i}a e Innovaci\'{o}n Productiva (Argentina), and Minist\'{e}rio da Ci\^{e}ncia, Tecnologia e Inova\c{c}\~{a}o (Brazil).
The scientific results reported in this article are based in part on observations made by the Chandra X-ray Observatory.
This research has made use of the VizieR catalogue access tool, CDS,
 Strasbourg, France.
The original description of the VizieR service was
published in A\&AS 143, 23.

%%%%%%%%%%%%%%%%%%%%%%%%%%%%%%%%%%%%%%%%%%%%%%%%%%



\bibliographystyle{mnras}

%%%%%%%%%%%%%%%%%%%%%%%%%%%%%%%%%%%%%%%%%%%%%%%%%%

%%%%%%%%%%%%%%%%% APPENDICES %%%%%%%%%%%%%%%%%%%%%

%\appendix


%%%%%%%%%%%%%%%%%%%%%%%%%%%%%%%%%%%%%%%%%%%%%%%%%%

% Don't change these lines
\bsp	% typesetting comment
\label{lastpage}
\end{document}